\documentclass[twocolumn]{aastex63}
\usepackage{soul}
\usepackage{graphicx}
\usepackage{enumerate}
\usepackage{hanging, booktabs}
\usepackage[caption=false]{subfig}
\usepackage{array}
\usepackage{times}
\usepackage{amsmath}
\usepackage{natbib} 
\usepackage{longtable}
\usepackage{afterpage}

\def\aj{{AJ}}                   
\def\araa{{ARA\&A}}          
\def\apj{{ApJ}}                 
\def\apjl{{ApJ}}                
\def\apjs{ {ApJS}}

\def\aap{ {A\&A}}                
          
\def\aaps{ {A\&AS}}

\def\mnras{ {MNRAS}}

\def\pasp{ {PASP}}

\def\arcsec{$^{\prime\prime}$}

\newcommand{\msun}{{$M_{\odot}$}}
\newcommand{\mstar}{{$M_{\star}$}}

\newcommand{\se}{s$^{-1}$ }
\newcommand{\degree}{$^{\circ}$} 
\newcommand{\gtsima}{$\; \buildrel > \over \sim \;$}
\newcommand{\ltsima}{$\; \buildrel < \over \sim \;$}
\newcommand{\prosima}{$\; \buildrel \propto \over \sim \;$}
\newcommand{\gsim}{\lower.5ex\hbox{\gtsima}}
\newcommand{\lsim}{\lower.5ex\hbox{\ltsima}}
\newcommand{\simgt}{\lower.5ex\hbox{\gtsima}}
\newcommand{\simlt}{\lower.5ex\hbox{\ltsima}}
\newcommand{\simpr}{\lower.5ex\hbox{\prosima}}
\newcommand{\es}{erg~s$^{-1}$}

\newcommand{\cxo}{\textit{Chandra}}

\newcommand{\lx}{$L_{\rm X}$}

\newcommand{\lm}{LMXB}
\newcommand{\hm}{HMXB}
\newcommand{\im}{IMXB}
\newcommand{\lms}{LMXBs}
\newcommand{\hms}{HMXBs}
\newcommand{\ims}{IMXBs}
\newcommand{\xlf}{XLF}
\newcommand{\lem}{L19}

\def\revise#1{{\textcolor{black}{#1}}}

\begin{document}
\title{Calibrating X-ray binary luminosity functions via optical reconnaissance I. The case of M83}
\author[0000-0002-4669-0209]{Qiana Hunt}
\affiliation{Department of Astronomy, University of Michigan, 1085 S University, Ann Arbor, MI 48109, USA}
\author{Elena Gallo}
\affiliation{Department of Astronomy, University of Michigan, 1085 S University, Ann Arbor, MI 48109, USA}
\author[0000-0003-0085-4623]{Rupali Chandar}
\affiliation{Department of Physics and Astronomy, University of Toledo, Toledo, OH 43606, USA}
\author{Paula Johns Mulia}
\affiliation{Department of Physics and Astronomy, University of Toledo, Toledo, OH 43606, USA}
\author{Angus Mok}
\affiliation{Department of Physics and Astronomy, University of Toledo, 2801 W Bancroft Street, Toledo, OH 43606, USA}
\author{Andrea Prestwich}
\affiliation{Harvard-Smithsonian Center for Astrophysics, 60 Garden Street, Cambridge, MA 02138, USA}
\author{Shengchen Liu}
\affiliation{Department of Astronomy, School of Physics, Peking University, Beijing 100871, China}

\begin{abstract}
Building on recent work by Chandar et al. (2020), we construct X-ray luminosity functions (XLFs) for different classes of X-ray binary (XRB) donors in the nearby star-forming galaxy M83 through a novel methodology: rather than classifying low- vs. high-mass XRBs based on the scaling of the number of X-ray sources with stellar mass and star formation rate, respectively, we utilize multi-band \textit{Hubble Space Telescope} imaging data to classify each \textit{Chandra}-detected compact X-ray source as a low-mass (i.e. donor mass $\simlt$ 3 \msun), high-mass (donor mass $\simgt$8 \msun) or intermediate-mass XRB based on either the location of its candidate counterpart on optical color-magnitude diagrams or the age of its host star cluster. In addition to the the standard (single and/or truncated) power-law functional shape, we approximate the resulting XLFs with a Schechter function. We identify a marginally significant (at the $1$-to-$2\sigma$ level) exponential downturn for the high-mass XRB XLF, at $\ell\simeq \revise{38.48 ^{+0.52}_{-0.33}}$ (in log CGS units). 
In contrast, the low- and intermediate-mass XRB XLFs, as well as the total XLF of M83, are formally consistent with sampling statistics from a single power-law. 
\revise{Our method suggests a non-negligible contribution from low- and possibly intermediate-mass XRBs to the total XRB XLF of M83, i.e. between 20 and 50\%, in broad agreement with X-ray based XLFs. More generally, we caution against considerable contamination from X-ray emitting supernova remnants to the published, X-ray based XLFs of M83, and possibly all actively star-forming galaxies.}
\keywords{X-rays: binaries -- X-rays: galaxies -- stars: luminosity function, mass function -- techniques: photometric}
\end{abstract}

\section{Introduction} \label{sec:intro}

In the absence of a bright active galactic nucleus (AGN), X-ray binaries (XRBs) dominate the point source X-ray emission of a galaxy, with only minor contributions from coronally active binaries, cataclysmic variables, unresolved sources and supernova remnants (\citealt{fabbiano06,boroson11}, and references therein). Owing to their different formation and evolutionary time scales, XRBs with low- and high-mass donors (hereafter referred to as \lms\ and \hms, where the former refers to donor masses below $\simlt 3$ \msun, whereas latter refers to donors in excess of $\simgt$8 \msun) can be expected to trace the integrated stellar content and recent star formation output of their host galaxy, respectively. Indeed, seminal work by \citet{grimm03}, based on \textit{ASCA} and early \cxo~\textit{X-ray Observatory} data for the Milky Way Galaxy and the Magellanic Clouds (see also \citealt{ranalli03}) first established a quantitative scaling between the integrated X-ray luminosity of \hms\ with the host galaxy star formation rate (SFR). In a follow-up study, \citet{gilfanov-sfr} crystallized the notion of a ``universal" \hm\ X-ray luminosity function (XLF) for star-forming galaxies, the normalization of which is proportional to the host galaxy SFR.  Along the same lines, using \cxo\ data for 11 nearby galaxies, \cite{gilfanov04} demonstrated that the total number of \lms\ and their integrated X-ray luminosity are proportional to the stellar mass (\mstar) budget of the host galaxy, thereby establishing a ``universal" XLF for \lms\ (see also \citealt{kim04}). 

Since these pioneering works, quantitative investigations of the scaling relations between the \hm\ and \lm\ XLFs and host galaxy properties have progressively sharpened, largely thanks to several Msec of sub-arcsec resolution X-ray imaging data accumulated by \cxo\ for tens of nearby galaxies. Updated XRB XLFs and their scaling relations are routinely employed for a variety of purposes. For example, the total X-ray luminosity is taken as a reliable SFR proxy in distant, star-forming galaxies \citep{mineo14}, and the expected total X-ray luminosity in XRBs \citep{lehmer10} can be used to argue in favor or against a low-luminosity AGN on a statistical basis.
Nevertheless, perhaps not surprisingly, a larger degree of complexity has also emerged from this wealth of data. 

There is reason to question the supposed universality of the XRB XLFs and their scaling relations. From a theoretical standpoint, large variations in the \lm\ XLFs are to be expected with stellar age, whereas metallicity effects are expected to drive variations in the \hm\ \xlf\ \citep{fragos13}. The notion that both may have sizable effects on the XRB XLFs is consistent with the claimed redshift evolution of \lx/\mstar\ and \lx/SFR, whereby the cosmic decline in mean stellar age and metallicity would be responsible for the inferred increase of both ratios with $z$ (\citealt{lehmer16,aird17}, and references therein). However, a quantitative assessment of the role of age and metallicity in shaping the XRB XLFs functional form is inevitably challenging, as deep, high-statistics XLFs have only been assembled for nearby galaxies spanning a relatively limited range in both (see, e.g., \citealt{irwin04,colbert04,kaaret11,prestwich13,plotkin14,bz16,tz16,wang16}). \\

There may also be issues with the overall approach to characterizing XRB XLFs. In an effort to isolate the \lm\ population, even the most comprehensive studies select massive elliptical galaxy samples, with large numbers of sources and negligible SFR levels, so as to ensure virtually zero \hm\ contamination. Conversely, \hm\ XLF studies focus on high specific SFR (sSFR, defined as the ratio SFR/\mstar) galaxies so as to minimize the contribution from \lms.  
Although practical, these selection strategies may affect the robustness of the inferred XLFs, or, at a minimum, pose a challenge to their claimed universality. For one, late-type galaxies with moderate sSFRs are excluded targets, as they are guaranteed to have a mixed \lm/\hm\ population. This, however, also means that we have \textit{no reliable knowledge of the \lm\ XLF in star-forming galaxies} (to this end, it is perhaps indicative that, when \citealt{mineo12} attempted to correct their inferred \hm\ XLF for possible \lms\ contamination using the XLF derived from elliptical galaxies, they obtained negative counts). Since the \lm\ contribution theoretically depends on mean stellar age, it is reasonable to expect that early and late-type galaxies may exhibit different correlations with stellar mass, as well as different spatial distributions. 

Lastly, there are indications that globular cluster (GC) specific frequency may also affect the shape and normalization of the \lm\ XLF \citep{sivakoff04,humphrey08,kim09,zhang10,peacock16}; this is not surprising, since field vs. GC \lms\ likely have different origins and evolutionary paths. If so, then the higher specific frequency of GCs in massive ellipticals could yield higher XLF normalizations than for \lm s in late-type galaxies. Additionally, metallicity can affect the XLF of GC XRBs, as red GCs may be more often associated with bright \lms\  \citep{jordan04,kim06,sivakoff07,peacock07,kim13,luan18}.\\

In a recent endeavor to properly characterize the XLFs of both \hms\ and \lms\ in late-type galaxies, \citet[][hereafter L19]{lehmer19} fit the XLFs of 38 nearby galaxies spanning a broad range of SFR and \mstar\ with a global model which fits simultaneously for the contributions from \hms, \lms\ and background sources using sub-galactic SFR and stellar mass maps (see also \citealt{lehmer17}, and references therein, for other examples of sub-galactic modeling studies). This novel and powerful approach reveals a smooth, progressive decline in the XLF normalization per unit SFR, accompanied by a decrease in normalization at the bright-end with increasing sSFR, i.e., as the dominant contribution to the XRB population shifts from \lms\ to \hms. The study also unveils further interesting subtleties, such as an intriguing flattening of the \hm\ \xlf\ between $10^{38}-10^{40}$ erg \se\, and a possible disagreement in the \lm\ \xlf\ slopes below $10^{38}$ and above $10^{39}$ erg \se\ compared to the results obtained for elliptical galaxies only \citep{zhang12}. This further emphasizes the need to move beyond a one-size-fits-all XLF modeling concept. \\

Motivated by the same kind of considerations, this Paper follows a very different approach to disentangling the \lm\ and \hm\ XLFs: by leveraging \textit{Hubble Space Telescope} (HST) multi-band imaging data, we \textit{directly classify the optical counterparts} to \cxo-detected point-like X-ray sources in the field of view of the target galaxy.  This technique, which was developed for and tested on the nearby spiral galaxy M101 by \citet[][hereafter C20]{chandar20}, hinges on the notion that, on average, HST imaging enables the direct detection of XRB donor stars down to a given distance-dependent mass limit (e.g., down to $\sim$3~M$_{\odot}$ at the distance of M101). By construction, this procedure does not rely on underlying assumptions about the relationship between a galaxy's XRB populations and their local environments; it also enables us, for the first time, to elucidate the role of intermediate-mass XRBs, i.e. XRBs with donors in the $\sim 3-8$ \msun\ range.

Here, we introduce a number of improvements upon C20 and apply our revised methodology to M83 (NGC~5236). M83 is a face-on (i~=~24\degree; \citealt{talbot79}) spiral galaxy with \mstar~$\sim2\times10^{10}$~\msun, a moderate SFR of $\approx 2.5$~\msun~yr$^{-1}$ (L19), and no significant AGN contribution in the nuclear region. At a distance of 4.66~Mpc \citep{saha06}, yielding a distance modulus of 28.32 and a physical scale of 1\arcsec $\approx$ 22~pc, it is closer than M101 (6.4~$\pm$~0.2~Mpc, with 1\arcsec~$\approx$~31~pc).
Furthermore, the Galactic absorption is low along the line of sight to M83 (N$\rm_{H} = 4\times10^{20} cm^{-2}$; \citealt{kalberla05}), making it ideal for an optical photometric study of X-ray source populations. \\

In this work, we present a fully classified catalog of X-ray sources in M83 that builds upon that published in L19. Each source is classified on a rigorous, source-by-source basis as either a low-, intermediate-, or high-mass XRB, a background galaxy, or a supernova remnant (SNR). We use these classifications to construct ``uncontaminated" XLFs for each XRB population.  Our main goal is to assess the shape of the XLFs and establish whether or not there is evidence for a statistically significant cut-off at the bright end, the presence of which has been widely debated (e.g. \citealt{zhang12}; \citealt{mineo12}; C20).
We are also interested in the normalization of each XLF, and how well it matches predictions from the global model presented by L19 and other studies. The rest of this work is organized as follows:  in \S\ref{sec:sourceclass}, we identify the optical counterparts to X-ray sources in M83, separating out contamination (AGN, quasars, and SNR, \S\ref{sec:nonxrb}) from the XRBs, and we estimate the masses of XRBs, which may appear as either an individual donor star or existing within a parent cluster; in \S\ref{sec:results-sd}, we investigates the spatial distribution of the classified XRBs; in \S\ref{sec:xlfs}, we present Schechter and power-law function fits to the the XRB XLFs and assess the presence of a downturn or cut-off; and finally in \S\ref{sec:discussion}, we compare the normalizations of our optical data-based XLFs with the literature (namely, L19).

\section{Source Classification} \label{sec:sourceclass}

\subsection{X-ray Source Catalog}\label{sec:xsources}
As our primary X-ray point-source list, we adopt the M83 catalog constructed from deep \cxo\ ACIS imaging data by L19, which examines a total of 38 galaxies. 
The L19 study includes a thorough estimate of the completeness of the detected X-ray point sources, which is crucial to our purposes.
The \cxo\ data were reduced following the methods detailed in \citet{lehmer17}: for each galaxy, the analysis was restricted to data sets with aim points within 5\arcmin\ of the nominal center position, ensuring a sharp point spread function for the nuclear regions which tend to be the most crowded. \revise{The source detection and parameter extraction were performed within 0.5-7 keV, where ACIS is best calibrated.}

\begin{figure}[t]
    \centering
    \includegraphics[width=0.9\linewidth]{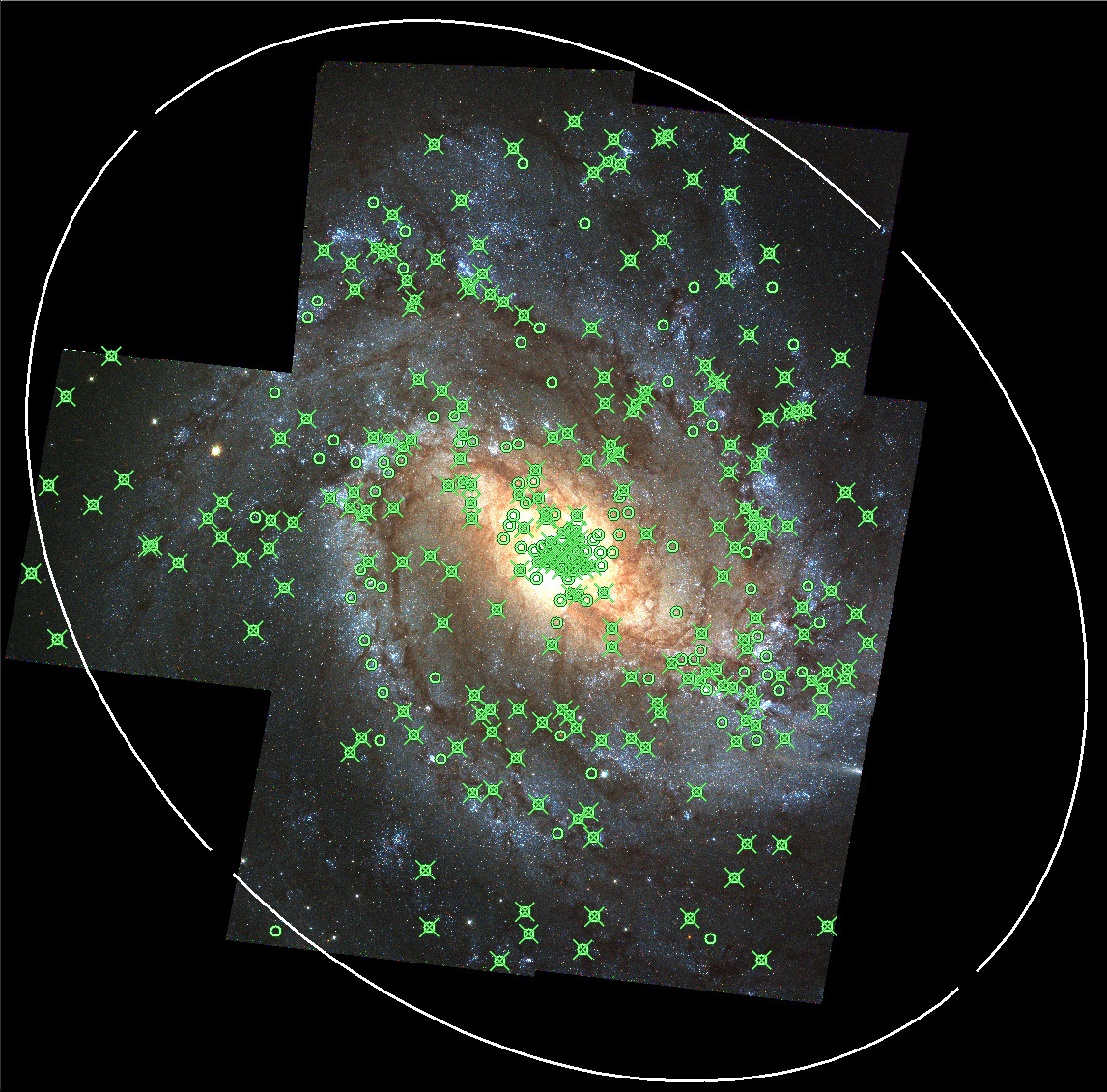}
    \figcaption{An optical image of M83 taken with by the WFC3 camera on HST \citep[][available at \texttt{https://archive.stsci.edu/prepds/m83mos}]{blair14}.  The $B$-band is shown in blue, $V$-band in green, and $I$-band in red.  The 7-pointing mosaic covers $\approx 43$ square arcminutes (75.2~Mpc$^2$)
The locations of all X-ray sources from the L19 catalog (adopted here and shown as green circles) and those from the version 2.0 release of the Chandra Source Catalog (green X's) are shown. The galactic footprint adopted by L19 tracing the {K$\rm_{s} \approx 20$ mag arcsec$^{-2}$} galactic surface brightness \citep[see][]{jarrett03} is outlined in white for comparison. \label{fig:mosaic}}
\end{figure}

Out of a total of 456 point-like sources\footnote{By comparison the \cxo\ Source Catalog Release 2 (CSC 2.0; \citealt{evans20}) returns 463 unique, significant X-ray sources within 12.\arcsec9 of the galaxy nominal center.} brighter than $10^{35}$ \es, we restrict our analysis to the 325 objects that fall within the M83 HST footprint, shown in Figure \ref{fig:mosaic}. For comparison, L19 restricts its XLF fitting to those 363 sources that are located within an ellipse that traces the $K_s\approx 20$ mag arcsec$^{-2}$ galactic surface brightness, outlined in white.\\ 

\citet{long14} also published a catalog of X-ray point sources in M83 based on a partial set of the data used in L19.  The main focus of their work was to detect a sample of SNRs using multi-wavelength observations.  In \S\ref{sec:nonxrb} we use classification information provided in the Long catalog to eliminate SNRs and some background AGN from our initial X-ray point source catalog.

\subsection{Combining X-ray and Optical Data}

HST observations of M83 were taken with the WFC3/UVIS instrument, spanning seven fields that each cover approximately 162\arcsec $\times$ 162\arcsec\ for a total mosaic area of $\sim$ 43 square arcminutes. All observations were obtained between August 2009 and September 2012 by R. O'Connell (Prop ID. 11360) and W. Blair (Prop ID. 12513), with exposure times ranging from $\sim 1.2-2.7$ ks for each image. Images were downloaded from the {Hubble Legacy Archive} (HLA\footnote{\texttt{http://hla.stsci.edu/}}). In general, BVI images are created using the F438W, F547M, and F814W filters. The central field, which includes the galaxy nucleus, uses the broader F555W $V$-band filter, rather than F547M. We also use U-band images (F336W) to help calculate cluster ages (see \S\ref{sec:clusterages}).  

Figure \ref{fig:mosaic} shows a BVI mosaic of all seven M83 HST fields (available on HLA; \citealt{blair14}). The mosaic combines the two different V-band filters used for the central field (F555W) and the 6 remaining fields (F547M) by scaling the F555W data to match the scaling on F547M. We utilize this single, cohesive V-band image for correcting the relative astrometry between \cxo\ and HST observations.  
6 X-ray sources whose HST counterparts are clearly background galaxies (i.e., morphology and color; see \S\ref{sec:nonxrb} for details) are identified as reference `AGN' for the astrometric correction. For these objects, we calculate a median relative positional offset of $\sim$0.\arcsec079 and 0.\arcsec229 along the x- and y- axes of the HST image respectively, with standard deviations of 0.\arcsec177 and 0.\arcsec182. The X-ray centroid positions of all sources in our sample are shifted by these offsets to the positions indicated on Figure \ref{fig:mosaic} in green. 

\subsection{Candidate Optical Counterparts}

Optical counterparts to XRBs in M83 can be either donor stars or host stellar clusters.
We use the IRAF DAOFIND task to detect all point-like sources, down to the faintest levels, on the composite V-band image.  
We perform aperture photometry with the IRAF PHOT task using a 3~pixel aperture radius for each detected source, with the local background level determined in an annulus with radii between 20 and 25 pixels.
Due to the rescaling of the F555W field for the creation of the mosaic, which may have introduced calibration errors, 
photometry of all detected sources is performed on individual images rather than on the mosaic.
Aperture corrections of -0.48~mag (V) and -0.61~mag (I) were determined by taking the median difference between the magnitude in 3 and 20 pixel apertures of several relatively bright, isolated stars with smooth radial profiles that flatten towards the background sky magnitude with increasing aperture radius. 

An additional correction term  of -0.06~mag is added to each filter to correct for the small amount of flux missing from a 20~pixel aperture (see Encircled Energy Fractions from 20 pixels to infinity in \citealt{deustua17}).
These instrumental magnitudes are converted to the VEGAMAG system by applying the zero-point magnitude for each filter as reported in Table 2 of \citet{deustua17}. 

Candidate optical counterparts to each X-ray source are initially selected by applying a proximity criterion to the X-ray source positions.  
We define 1- and 2-$\sigma$ positional uncertainty radii for each source by adding in quadrature the standard deviation in the Chandra-HST positional offsets and the X-ray positional uncertainty (see Figure \ref{fig:mosaics}). The latter depends sensitively on the X-ray source distance from the observation aim-point, as well as the number of counts. We adopt Equations 14 and 12 in \citet{kim07} to calculate the 68\% and 95\%\ confidence positional uncertainty for each source, as a function of total counts (C) and off-axis angle (OAA) -- both of which are available in the L19 catalog. An additional uncertainty term is added to the 1- and 2-$\sigma$ radii calculated above, due to the slight rotations between the fields and the mosaic. This is an improvement on the method used in C20, where the 1- and 2-$\sigma$ positional uncertainties in M101 were assume to be circles with a radius of $0.3$\arcsec\ and $0.6$\arcsec\ for each X-ray point source.
A final correction is made to the absolute optical magnitudes of each source to account for foreground extinction. Using  the relation ${\rm N_{H} (cm^{-2}) = (2.21~\pm~0.09)\times10^{21} A_{V}}$ \citep{guver09} with the known Galactic absorption towards M83 \citep{long14}, we find an extinction of ${\rm A_{V}\approx0.174}$ mag, corresponding to a reddening of ${\rm E(B-V)\approx0.054~mag}$\ \citep{mathis90}. We do not account for extinction intrinsic to each source, though we employ a confidence flag scheme (see Table \ref{tab:allsources}) to indicate sources that may be particularly susceptible to the effects of reddening and obscuration within M83. \revise{We test the impact of this assumption on our results in \S\ref{sec:donormass}.}

\subsection{Non-X-ray Binary Sources}\label{sec:nonxrb}

\begin{figure}[t]
    \centering
    \includegraphics[width=\linewidth]{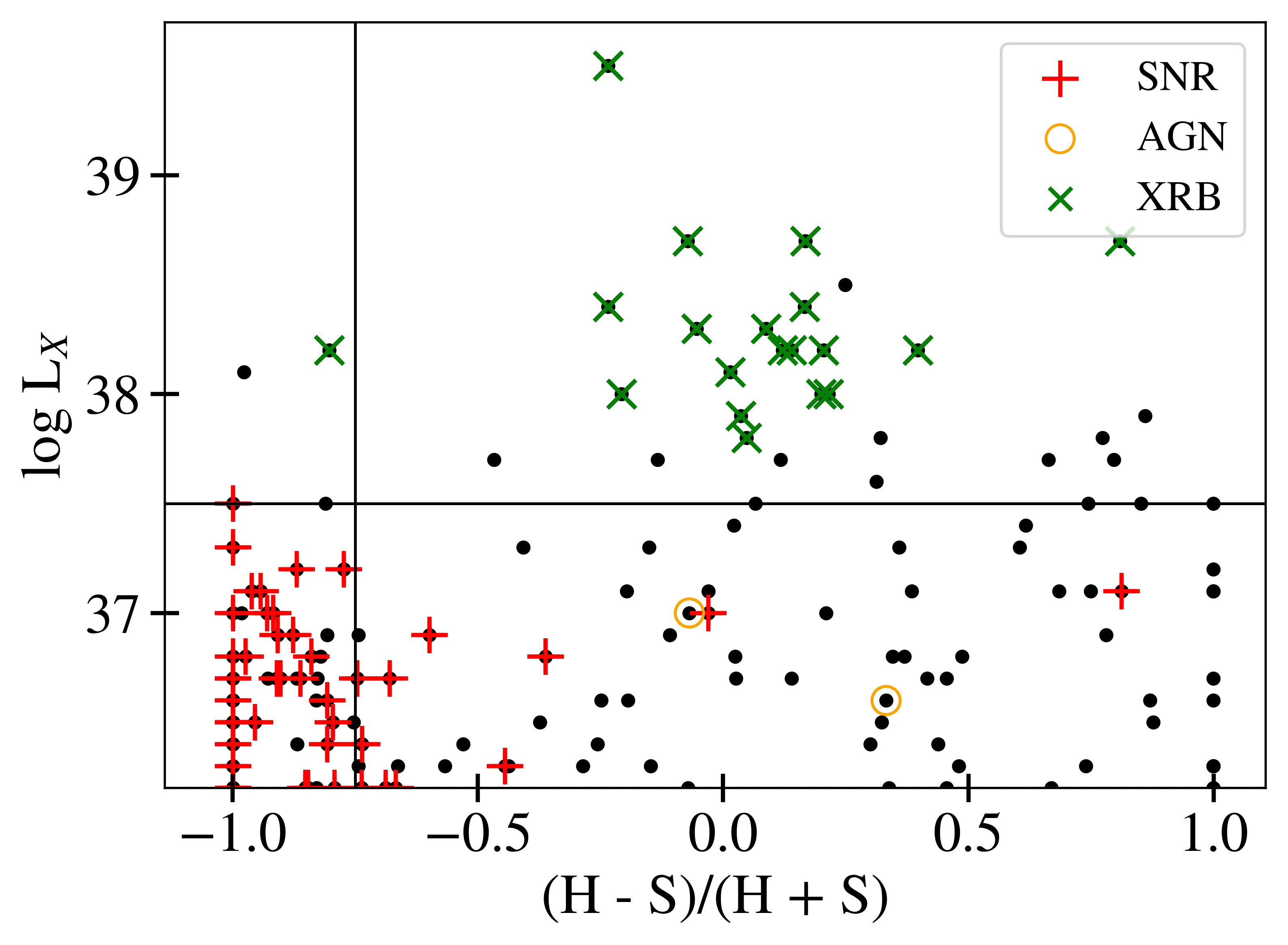}
    \figcaption{The measured X-ray luminosity vs. hardness ratio of sources in M83, with the subset classified as SNRs (red +'s), AGN (yellow circles), and XRBs (green X's) by \citet{long14} indicated. SNRs occupy a softer portion of this parameter space that does not overlap with AGN or XRBs. The hardness ratio is defined as the difference between the hard band (2-7~keV; H) and the soft band (0.5-1.2~keV; S) \cxo\ counts over their sum. We define a selection criteria that cuts sources with X-ray log luminosities below 37.5 and hardness ratios less than -$0.75$, the limit softer than which 60\% of the sources are SNRs, to minimize contamination from unclassified SNRs in our sample.\label{fig:HR}}
\end{figure}

Contributions from stellar sources such as coronally active binaries and cataclysmic variables are completely negligible above $L_X\simeq 10^{36}$ \es\ \citep{boroson11}, the completeness limit of our sample (L19). Hence, the main sources of contamination are background AGN and quasars, and bright SNRs within the host galaxy. We address them in turn below. Our approach differs substantially from all other XLF investigations in that we aim to directly identify and reject all contaminants, whereas published works almost exclusively correct for AGN contamination statistically, using the known Cosmic X-ray Background $\log N-\log S$. \\

In an extended, multi-wavelength spectral and temporal analysis of M83's X-ray point-source population, \citet{long14} classified a significant fraction of the point source population as SNR. These were classified based on variability, spectral hardness, [S II]:H$\alpha$ line ratios or strong [O III] emission, and cross-referencing with earlier SNR catalogs using \cxo, \textit{XMM-Newton}, Magellan, the Australia Telescope Compact Array, and the sites of historical supernovae \citep{wood74, soria03, maddox06, dopita10, blair12, ducci13}. 

Long's investigation provides us with the rare opportunity to construct a set of X-ray based diagnostic criteria that can be used to reject contaminants that have not been previously classified, and that are typically ignored in other studies. Out of the 87 X-ray sources identified by \citet{long14} as SNRs or SNR candidates, 76 are included in the HST footprint. Of these, 55 have available X-ray soft and hard counts ($S$ and $H$, corresponding to the 2-7 keV and 0.5-1.2 keV bands, respectively) in the CSC R2. We obtain a hardness ratio, defined as the difference between H and S over the total counts in both, for each source where available. We plot the measured X-ray luminosity, \lx\, against the hardness ratio in Figure \ref{fig:HR}. \revise{The value of \lx\ for each source is taken from L19, in which luminosities are obtained from 0.5-8 keV X-ray fluxes. This energy range allows for a straightforward comparison with (most of) the literature (see L19 for details).} The \citet{long14} SNR (red $+$'s), AGN (orange circles) and XRB (green X's) classifications are indicated where given. 

We find that the majority of SNRs belong to a distinct parameter space that is well-separated from other X-ray sources. Based on this phenomenological approach, we adopt minimum \lx\ and HR cuts to automatically select out candidate SNRs in our catalog. Our \lx\ cut is set to the highest \lx\ of the Long SNRs in our sample, or $\ell \le 37.5$, where (hereafter) $\ell$ represents logarithmic X-ray luminosities in units of \es.
The minimum HR is chosen such that 60\% of all sources softer than this limit are Long-classified SNRs, which corresponds to HR$~\le-0.75$. Excluding (76) X-ray sources classified by \citet{long14} as candidate SNRs, an additional 27 sources meet our criteria. All 27 X-ray sources with these properties are rejected from our catalog. We assess the impact of this on the XLFs in \S\ref{sec:xlfcomp} and the Appendix. \\

After removing SNRs, the remaining population of contaminants consist of background AGN and quasars.
The rest may be identified through catalog cross-referencing and an analysis of optical properties. \citet{long14} classify several foreground and background contaminants based on X-ray hardness, X-ray-to-optical flux ratios, color, and prior catalogs. We classify additional contaminants on the basis of their morphology. 
Background galaxies can typically be identified by their distinct morphologies, including features such as nuclei, spiral arms and/or disks, with the exception of distant quasars, which appear in optical images as primarily red sources with extended radial profiles.  
In total, we remove from our sample 3 visually identifiable background AGN and 2 sources identified as AGN in \citet{long14}. We also remove \revise{2} candidate quasars.

\subsection{Classification of XRBs Based on Parent Cluster Age}\label{sec:clusterages}

We expect to find some XRBs in our sample still occupying their parent star clusters. In early-type galaxies, between 25-70\% of LMXBs are found in ancient globular clusters \citep{angelini01, kundu02, jordan04, kundu07, humphrey08, peacock16}.  In spiral and star-forming galaxies, however, the fraction of LMXBs found in globular clusters remains quite uncertain, albeit significantly lower: the dwarf starburst NGC~4449 and the spiral galaxy M101 each has only a single LMXB in a globular cluster (\citealt{rangelov11}; C20).  The fraction of HMXBs in spirals which still reside in their parent clusters is also poorly known but higher, with $\approx 15$\% of XRBs in M101 (C20) and $\approx25$\% in the Antennae \citep{rangelov12} found in clusters younger than a few 100~Myr.

Because of their high stellar density, the donor stars feeding XRBs that reside within compact stellar clusters cannot be identified individually at the distance of M83. However, the ages of parent clusters can be used as a proxy for estimating the masses of the donor stars in XRBs, since the most massive surviving stars within a cluster are the most dynamically active, and hence the most likely to form tight binaries. High mass stars ($\geq8~M_{\odot}$) have hydrogen burning lifetimes of only $\sim10$~Myr, and intermediate-mass ($\gtrsim3~M_{\odot}$) stars have lifetimes of $\sim400$~Myr. This means that clusters older than $\sim400$~Myr only contain stars less massive than $3~M_{\odot}$ and may only host \lms, while clusters younger than $\sim$10~Myr are likely to host HMXBs.  We assume that clusters with ages between 10 and 400~Myr host IMXBs since the most massive stars remaining in clusters in this age range (3-8 $M_{\odot}$) have intermediate mass.

\begin{figure}[t]
    \centering
    \includegraphics[width=\linewidth]{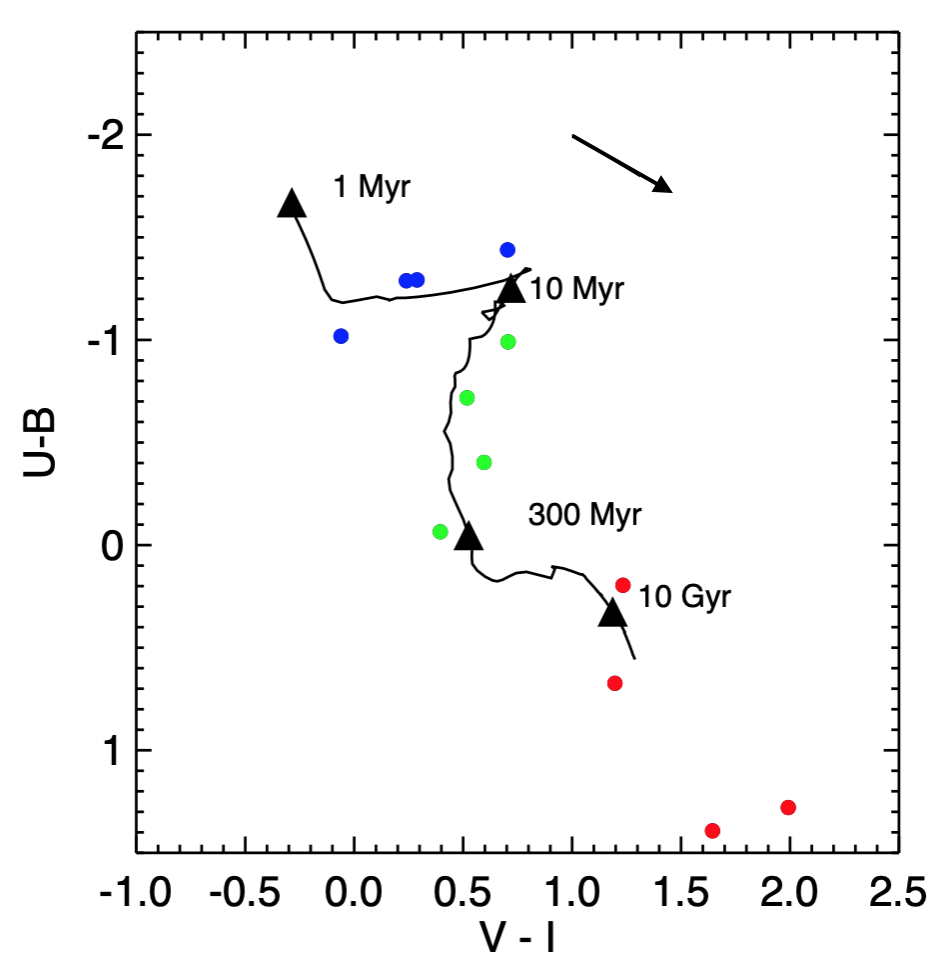}
    \figcaption{Measured $U-B$ vs $V-I$ colors of XRB host clusters compared with predictions for the color evolution of clusters from the \citet{bruzual03} models.  Different model ages are marked by black triangles, with blue points indicating \hms, green indicating \ims, and red indicating \lms, moving from top left to bottom right.  The arrow represents the direction of reddening expected from a Milky Way-type extinction curve with $A_V=1$. \label{fig:cluster}}
\end{figure}

To identify possible clusters among our candidates, we compare our optical matches for each XRB to a catalog of M83 clusters previously published by \citet{chandar14}. This catalog selected clusters to be broader than the point spread function (PSF) based on the FWHM of the radial profile, as well as the Concentration Index, defined as the difference in magnitudes measured for a 1 and 3 pixel aperture radius (see \citealt{chandar14} for details about selection).  We find that a total of \revise{12} ($\sim5$\% of the total) XRBs in M83 are found within a compact stellar cluster.  We compare the colors measured for the XRB-cluster hosts with those predicted by the \citet{bruzual03} stellar evolution models at solar metallicity in Figure~\ref{fig:cluster}.  These models start at 1~Myr (upper left) and go through 13~Gyr (lower-right), with key ages marked along the model track.  The arrow indicates the direction the colors of clusters would move due to reddening by dust.  XRB cluster hosts have a range of colors and, hence, ages.

While the color-color diagram provides a good visual comparison of cluster colors with model predictions, the age of each cluster is estimated using a spectral energy distribution fitting method, where we fit for the best combination of age and reddening, as described in \cite{chandar14}.  Magnitudes measured in the UBVI and H$\alpha$ filters for each cluster are fit to predictions from \citet{bruzual03} using a standard $\chi^2$ minimization; see \citet{chandar14} for details. The best fit age for each cluster is recorded in Table~\ref{tab:allsources} and used to classify each XRB: HMXBs have parent clusters $<10$~Myr, IMXBs have parent clusters with best fit ages between $10-400$~Myr and LMXBs have parent cluster ages $>400$~Myr.

\begin{figure*}
\centering
\includegraphics[width=0.45\textwidth]{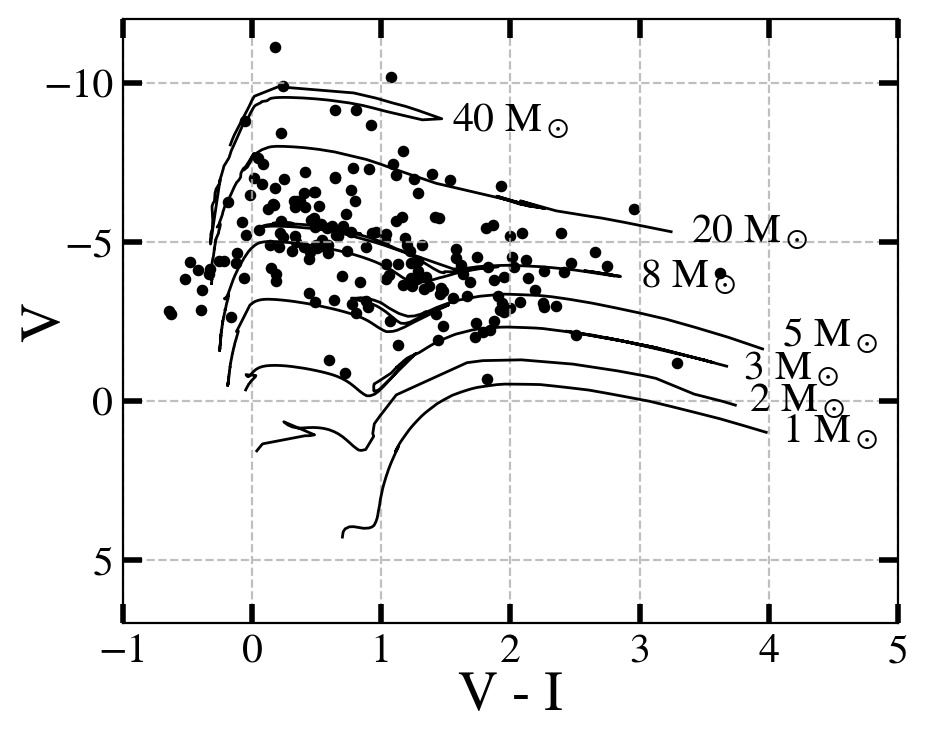}
\includegraphics[width=0.45\textwidth]{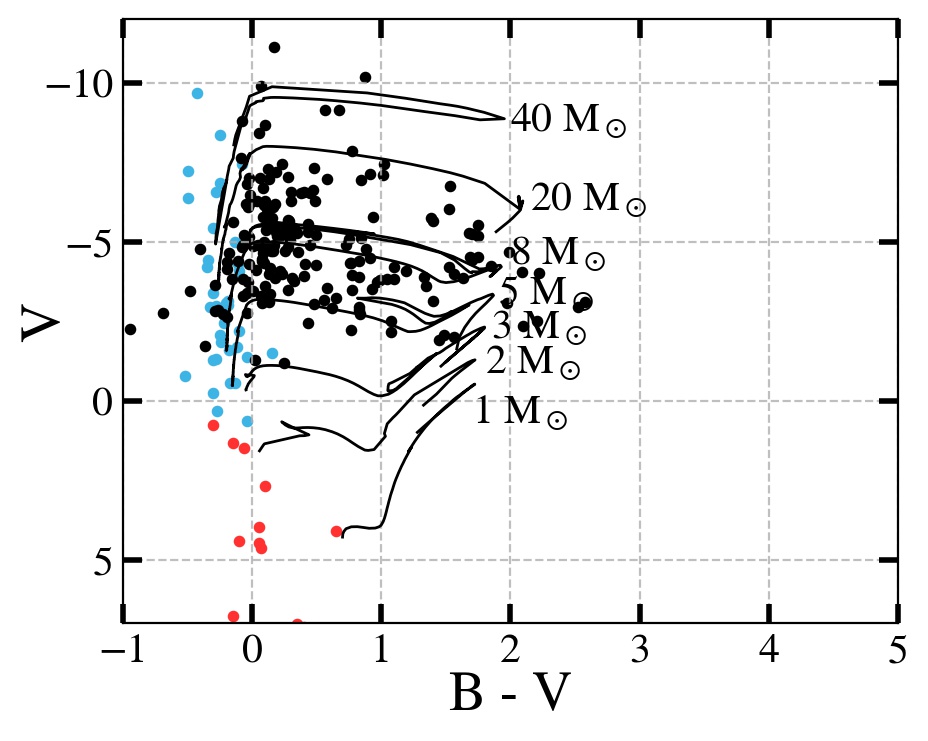}
\caption{Left: $V-I$ vs. $V$ CMD of XRB donor star candidates identified in M83 (black points). These are compared with theoretical evolutionary tracks modeled at solar metallicity  \citep{bertelli94, girardi10, marigo17}. Overall, donor stars in M83 appear to be detectable with the HST down to $\simgt 3$ \msun. Right: $B-V$ vs. $V$ CMD of candidate XRB donor stars. In addition, all X-ray bright Milky Way \hms\ and \lms\ having measured $B$ and $V$ magnitudes are shown as blue and red points, respectively, as identified by \cite{liu-h} and \cite{liu-l}. \label{fig:CMD}}
\end{figure*}

\subsection{Classification of XRBs Based on Donor Star Mass}\label{sec:donormass}

The majority of X-ray sources contain at least one optical point source within their 2-$\sigma$ radius, with several coincident with multiple candidates. The most likely XRB donor is chosen on a case-by-case basis. In general, priority is given to brighter sources that fall within or closest to the 1-$\sigma$ radius. In most cases, it is not necessary to identify the exact donor of an XRB with several bright candidates, so long as the candidates fall within the same mass regime. \revise{For cases in which multiple sources are detected over a range of possible donor masses, intermediate-mass sources are deprioritized, since few IMXBs have been identified in the Milky Way compared to high- and low-mass counterparts. When given the option between high-mass and low-mass potential counterparts, high-mass stars are prioritized as the most likely donor, given the relative rarity of both high-mass stars and XRBs within galaxies (see \S\ref{sec:misclass}).}

We directly estimate the masses of bright donor candidates by comparing them to the theoretical evolutionary mass tracks for solar metallicity stars from the Padova models on a color-magnitude diagram (CMD, Figure \ref{fig:CMD}), using the mass limits defined in \S\ref{sec:intro} (\lm\ $\leq$ 3 \msun, \hm\ $\geq$ 8 \msun).
C20 find that archival HST images of M101 are deep enough to see stars down to 3 \msun\ at a distance of 6.4~$\pm$~0.2~Mpc. Consistent with this, we find that the majority of the M83 XRB donors lie above the 3 \msun\ line, with a few sources falling below. This suggests that, indeed, we are able to detect sources down to the minimum threshold required to identify the donor stars of HMXBs within M83. On the other hand, X-ray sources that lack an optical counterpart likely have stellar components that fall beneath the observable magnitude threshold, suggesting the system is a \lm. 

The left and right panels of Figure \ref{fig:CMD} compare the V-I colors and B-V colors of the XRBs. We note that we are unable to extract B-band magnitudes from all of the point sources in our sample since the observations are not as deep in the B-band as they are in the V-band. Nevertheless, it is the absolute $V$-band luminosity that is most important for estimating the masses of the donor stars.

\revise{In order to investigate the potential effects of extinction of the donor stars in our X-ray binaries, we can compare the measured $U-B$ vs. $V-I$ colors of donor stars with predicted stellar tracks, following the procedure in \cite{kim17}.  Reddening will move the observed colors off the model tracks and allow for an estimate of $E(B-V)$.  This method requires photometric errors to be less than $\approx0.1$~mag in each band, so it is only applicable to a subsection of donor stars.  The colors of the donor stars that satisfy this requirement indicate low typical $E(B-V)$ values of $\sim0.1$ to $0.2$~mag, which is reasonable given that M83 is oriented nearly face-on.  This level of extinction would only affect the classification of a few IMXBs, in the sense that they would just exceed the $8~M_{\odot}$ theshold of the model tracks, but otherwise does not have much impact on our classifications. }\\

In total, \revise{214} of the 325 X-ray sources that fall within the HST footprint are classified as XRBs using the methods described, \revise{12} of which exist within compact stellar clusters. We present these sources, their positions, X-ray luminosity, optical colors, and source classification in Table~\ref{tab:allsources} and Figure~\ref{fig:mosaics}.
For these X-ray sources, we find \revise{30} are \lms, \revise{120} are \hms, and \revise{64} fall in the intermediate mass range between the two limits. For each of the panels in Figure~\ref{fig:mosaics}, the 1- and 2-$\sigma$ positional uncertainties are shown as yellow concentric circles, each detected optical source within the 2-$\sigma$ radius is highlighted by dashed yellow circles, and the chosen donor is indicated with bold red circles. In both Table~\ref{tab:allsources} and Figure~\ref{fig:mosaics}, italicized SNRs are those classified using our HR-\lx\ criterion described in \S\ref{sec:nonxrb} (as opposed to those identified directly in \citealt{long14}), italicized XRBs are those associated with clusters, and classifications in parentheses are objects with uncertain ``candidate" classifications, as reported in \citet{long14} or as determined by our methods.

\subsection{Assessing Misclassifications}\label{sec:misclass} 

While careful consideration is given to identifying LMXBs and HMXBs from each other and from non-XRBs, misclassifications are still possible. Here, we discuss possible sources of misclassification, our methods for mitigating these occurrences, and the impact their inclusion could have on the final XLFs.

\begin{itemize}
    \item A background AGN/quasar that is severely obscured by an optically thick portion of M83 could mimic an X-ray source with no detectable optical counterpart (LMXB). This type of misclassification will preferentially affect the inner and disk regions in a relatively small area of the total coverage.  We inspect the color mosaic image and estimate the dust obscured area within which we cannot see background galaxies to be roughly 20,000 arcsec$^{2}$. Scaling the detected rate of 1,000 sources per square degree in the Chandra Deep Field down to similar flux levels as used here \citep{luo17}, we expect 1-2 background galaxies to fall within these optically thick regions and hence be misclassified as a LMXB. The number of background galaxies expected across the area covered by the full mosaic (155,402 arcsec$^{2}$) is 11-12; we identify \revise{7}. One of the galaxies in our sample (L19X178) is totally obscured. In all, our observations roughly match expectations.
    
    \item Distant quasars appear as red, point sources in optical images and could be mistaken for a red giant donor star in a HMXB system. In fact, we identify \revise{2} optical counterparts to X-ray point sources that we classify as candidate quasars, based on their extended radial profiles and red colors. Overall, the space density of quasars on the sky is quite low, with only $\sim100$ expected per square degree with X-ray fluxes in our catalog. This suggests we should expect 1-2 quasars in our field of view (155,402 arcsec$^{2}$), consistent with our results.
    
    \item A star in a dusty region can experience partial or total extinction, resulting in a lower mass estimate.  Potentially, this could lead to HMXBs misidentified as IMXBs, particularly those that are near the 8 $M_{\odot}$ track. We addressed the possible effect of extinction in \S\ref{sec:donormass}.
    \smallskip 
    
    On the other hand, it is highly unlikely that a HMXB would be misidentified as a LMXB based on extinction in M83, because of the significant level required, which is not supported by a visual inspection of the optical color images. 
    
    \item A high mass star may happen to lie coincident with an LMXB, causing the LMXB to be misidentified as an HMXB. 
    Massive stars are rare compared to lower mass stars within galaxies and are highly concentrated, spatially, to regions of active star formation, such as the spiral arms (as we find for M83 in \S\ref{sec:results-sd}). Similarly, XRBs themselves are uncommon. Statistical arguments therefore suggest that the chance superposition of these two relatively rare phenomena is unlikely. \smallskip
        
    As a final precaution, we compare our maps of the XRB populations to the stellar mass and SFR maps of M83 published by L19. Since LMXBs and HMXBs are tracers of stellar mass and sSFR respectively, we expect a correlation between the locations of these populations and peaks in the stellar mass and sSFR maps. We examine these in \S\ref{sec:results-sd}. 
    
    \item A parent cluster might be misidentified as a single donor star, leading to an improper mass estimate using its $V-I$ color and $V$ magnitude rather than the age of the cluster.
    We expect very few, if any, misclassifications of this type, since clusters at the distance of M83 are more extended than the PSF.  To minimize mis-identifications, we cross-reference our sources with the published M83 cluster catalog published by \citet{chandar14}.  
    
    \item An LMXB could have flared at the time of observation, causing its disk luminosity and color to mimic that of a higher mass donor star.  The probability of this occurring is statistically negligible, at the level of 1 object per Milky Way stellar mass or so (well above the inferred stellar mass content of M83). \revise{The only persistent Milky Way analog is the black hole XRB GRS~1915+105. Less than a handful of other Galactic systems, mainly long-period neutron stars, have comparable luminosities, but the associated outburst duty cycles make them also statistically negligible.}\smallskip
    
    \revise{On the other hand, an \lm\ that evolved from an \im\ progenitor may appear bright enough to be mistaken for a higher-mass binary. We discuss this possibility at length in \S\ref{sec:donors}.}
\end{itemize}

\section{X-ray source Spatial Distributions}\label{sec:results-sd}

In general, owing to the the short lifetimes of the donors, HMXBs trace regions of recent star formation, whereas LMXBs trace the integrated stellar mass content. Almost all previous works have taken a statistical approach to classifying and studying populations of HMXBs and LMXBs, typically based on their location relative to different galactic structures (e.g., bulge, disk or outer region, as identified by, e.g., \citealt{mineo12}) or based on the SFR or stellar mass at their location (L19).  However, there are dynamical processes that can impart high space motions to XRBs, and thereby move them away from their sites of formation, and we would not expect a perfect spatial correlation, in any case. This means that statistical and spatially-based classifications are likely to have at least a few erroneous classifications of individual sources.  In this Section, we study the locations of HMXBs, IMXBs, and LMXBs based on our source-by-source classification method.

\begin{figure}[t]
    \centering
    \includegraphics[width=0.8\linewidth]{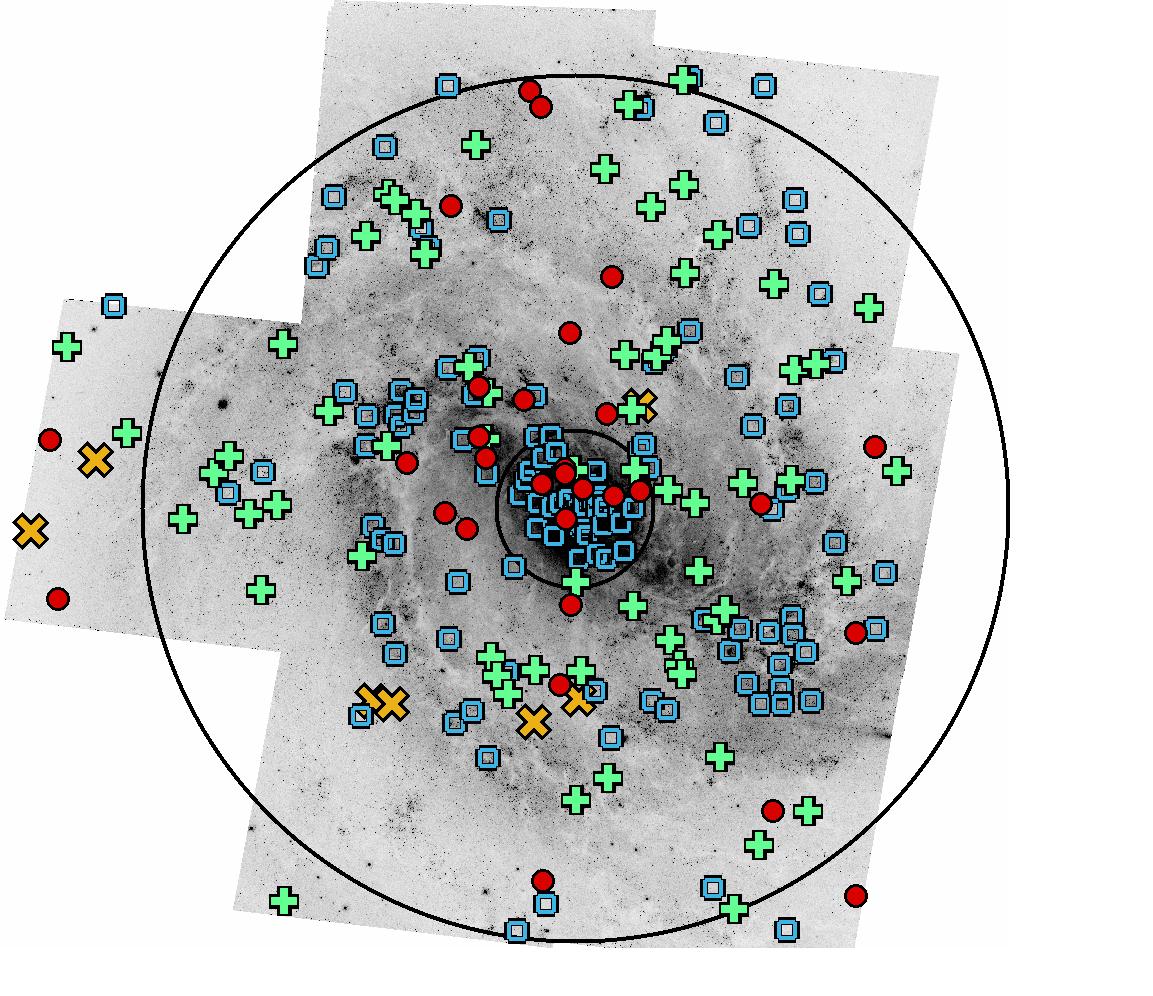}
    \vspace{0.2cm}
    \caption{Spatial distribution of LMXBs (red points), IMXBs (green crosses), and HMXBs (blue squares) in M83, as well as background galaxies (orange X's). Black outlines encircling the bulge and inner disk regions at $\sim$ 0.\arcmin66 and 3.\arcmin66 follow the prescription by \cite{mineo12}, with the bulge radius from \cite{dottori08}.}
    \label{fig:sd}
\end{figure}
 
\begin{figure}
\centering
   \subfloat{Stellar Mass (a) }\vspace{0.1cm}{
        \includegraphics[width=0.37\textwidth]{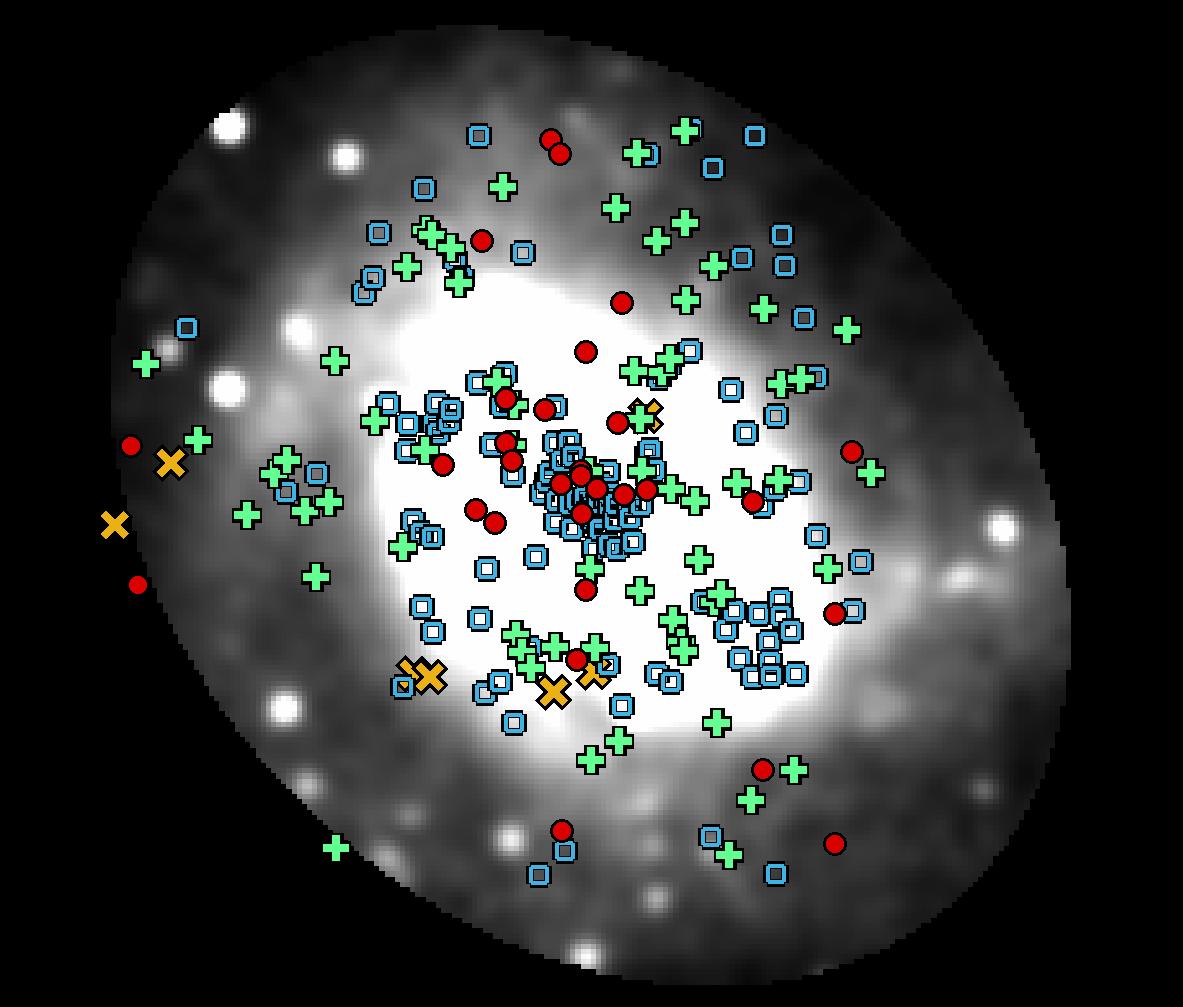}
        }
   \subfloat{Star Formation Rate (b)}\vspace{0.1cm}{
    \includegraphics[width=0.37\textwidth]{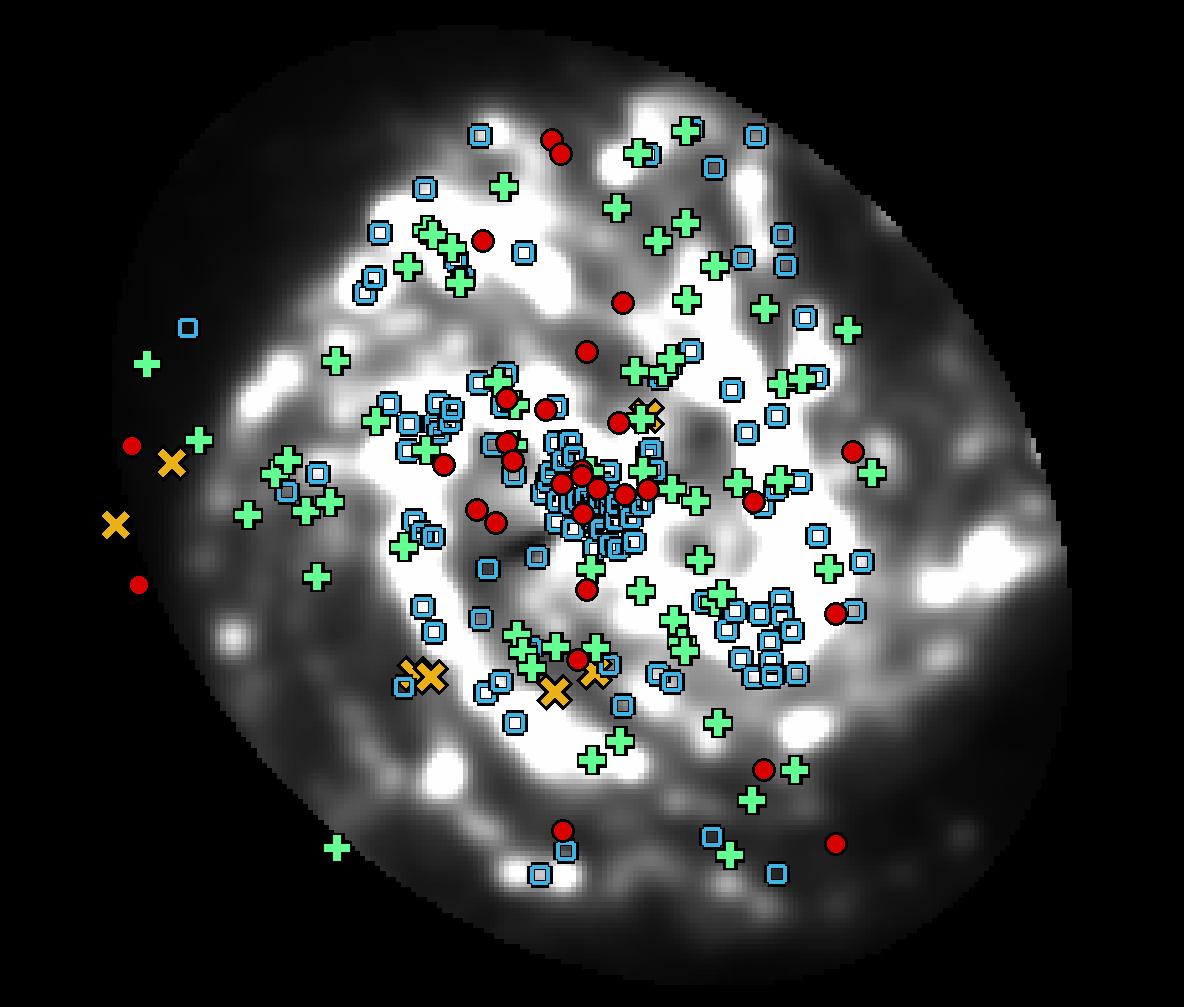}
        }
   \subfloat{Specific Star Formation Rate (c)}\vspace{0.1cm}{
    \includegraphics[width=0.37\textwidth]{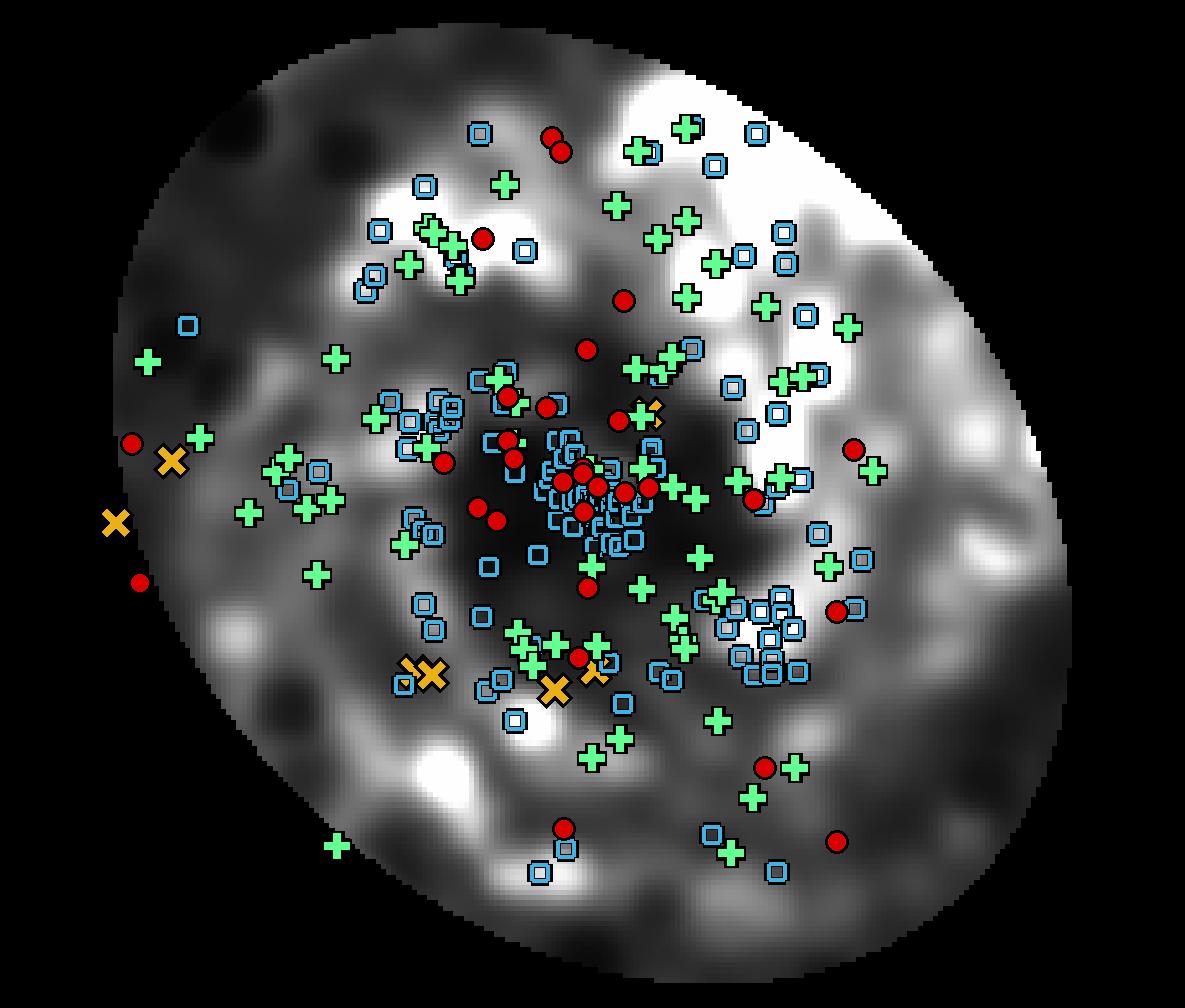}
        }
\caption{Overlays of LMXBs (red points), IMXBs (green crosses), HMXBs (blue squares), and background galaxies (orange X's) onto the (a) stellar mass, (b) star formation rate, and (c) specific star formation rate maps for M83 generated by L19. All three maps are shown with a linear color scale. \label{fig:lehmers}}
\end{figure}

Figure \ref{fig:sd} shows the spatial distribution of each class of XRB: \hms\ (blue squares), IMXBs (green crosses), and \lms\ (red points). We also find \revise{7} AGN/quasars (orange X's). The classifications are over-plotted on three different maps of M83 constructed by L19 in Figure \ref{fig:lehmers}: stellar mass ($M_\star$; top), SFR (middle), and specific star formation rate (sSFR$=$SFR$/M_\star$; bottom).   
These maps show both a high stellar mass \emph{and} a high SFR in the central region, higher SFRs in the spiral arms, and significantly higher sSFR in the arms compared to the inter-arm regions.

In the central 0.\arcmin66 or 894~pc of M83, there are a total of \revise{46} XRBs; of these, \revise{36} are HMXBs and \revise{7} are LMXBs. This high fraction of HMXBs is perhaps not surprising, since M83 has a central starburst.  However, it is different than the central region of M101, which is strongly dominated by LMXBs, even though it is a later-type galaxy with a smaller bulge.

Most HMXBs outside of the central region are found in regions of high sSFR.  We find a few in `dark' regions in the middle and bottom panels of Figure~\ref{fig:lehmers}; these may be sources with high space motions that have moved from their birth-sites. HMXBs also tend to be preferentially found in the spiral arms, with a fairly `clumpy,' rather than even, distribution.

LMXBs appear fairly centrally concentrated as expected from an old spheroidal (bulge/halo) population.  There are, however, sources distributed fairly evenly throughout M83, which may come from an old disk population.  There are also a few LMXBs further away from the center and located in regions of high sSFR, i.e. where the SFR strongly dominates over the stellar mass.  We checked the local background in these regions; extinction appears to be lower than closer to the center of M83, and the background level is sufficiently low that we would easily be able to detect donor stars down to 3 \msun.  These sources would be misclassified as \hms\ in studies that use a spatial approach, since they fall within a region of high sSFR (L19) and also within the `disk' region, as defined by \cite{mineo12}, typically believed to be dominated by \hms.
Overall, the spatial distribution of the XRB population is fairly mixed: whereas \hms\ and (to a much lesser extent) \ims\ seem to track higher SFR regions than \lms, \lms\ can be found in the bulge as well as the outer disk, and the bulge itself is home to a large \hm\ population.

%
\section{X-ray Luminosity Functions}\label{sec:xlfs}
A key goal of characterizing the XLF(s) of XRBs is to ascertain whether a statistically significant downturn exists at high luminosities. Theoretically, such a ``cutoff" is expected near the Eddington luminosity for stellar-mass compact objects, but observations have yet to confirm this prediction with high confidence. 
Assessing the presence and robustness of such a downturn is inherently dependent upon the chosen functional shape of the XLFs. Virtually all investigations of XRB XLFs perform either single power-law (PL) or broken-power law (BPL) fits to the differential and/or cumulative XLFs, often with an assumed cutoff luminosity. To first order, most studies conclude that a single PL with a high-energy cutoff at or above the maximum measured XRB luminosity adequately describes the shape of the luminosity distribution of \hms\ in highly star forming galaxies \citep{mineo12}. For \lms, the XLF is modeled with either a single or a broken power-law \citep{kim04,kim09,zhang12,lehmer19}, plus a high-luminosity cutoff.

There is, however, little physical motivation for adopting a BPL shape for the \lm\ XLF; this approach dates back to the seminal work by \cite{gilfanov04}, who first noted that the XLF of \lms\ in external galaxies ``is consistent with a power law with differential slope of $\simeq 1$ at low luminosities, gradually steepens above $\ell=37.0 - 37.5$ and has a rather abrupt cut-off at $\ell=39.0 - 39.5$." In later works, the best-fit values of the break and cutoff luminosities fall near $\ell\simeq 38$ and $\ell\simeq 40$, respectively (e.g., \citealt{mineo12}, L19). However, the lack of a unified approach (e.g., fitting cumulative vs. differential and/or binned vs un-binned distributions; fixing vs. fitting for the cutoff luminosity) makes it somewhat difficult to compare results across different studies. Furthermore, whether the presence of a break and/or a cutoff in the XLFs is required by the data with high statistical confidence remains an open, key question, whose answer is again intertwined with the choice of the XFL functional shape.  \\

As shown by \cite{mok19} in their thorough exploration of the mass function of young star clusters (see also C20, and references therein), {methods that bin the differential distribution in luminosity (or mass) intervals result in stable fits for the power-law indices}, while {fits to the un-binned distributions give the most robust detection of any downturn at higher luminosities}.

Guided by the above considerations, we approximate the shape of the XLF for XRBs in M83 with two functional shapes: a single PL, and a Schechter function \citep{schechter76}. We fit the X-ray luminosity distribution of the (a) composite sample (i.e., all XRBs), as well as (b)~\hms, (c)~\ims, and (d)~\lms, separately. \revise{Whereas the compact source fluxes were originally extracted over the 0.5-7 keV energy range, for the purpose of XLF fitting, L19 converts the source fluxes to a 0.5-8 keV range, as appropriate, i.e. using either a template spectral shape or the actual source spectrum, depending on the number of counts. This was meant to facilitate a comparison with the literature; we adhere to the same choice in this Paper}. 

To assess the presence of a truncation -- here defined as a luminosity above which no sources exist --  we adapt the methodology developed by \cite{mspecfit} to investigate the shape of the mass function of giant molecular clouds.
To account for the presence of a maximum luminosity value ($L_c$) in the distribution, we approximate the cumulative distribution as:
\begin{equation}
N(>L)= N_C \left[\left(\frac{L}{L_c}\right)^{\beta+1}-1\right],
\end{equation}
where $N_C$ is the number of XRBs more luminous than $2^{1/(\beta+1)}L_c$, at which point the distribution shows a significant deviation from a single power law of index $(\beta+1)$. In the case where $N_{0} \simeq 1$, there is no significant deviation, and the distribution is consistent with sampling from a single power law. With this formalism, the cumulative mass distribution below $L_c$ is proportional to $({L}/{L_c})^{\beta}$. \\

The differential Schechter luminosity distribution is proportional to $(L/L_{\star})$, where $L_{\star}$ | known as the Schechter ``knee" | corresponds to a characteristic luminosity above which the distribution declines exponentially, as follows:
\begin{equation}
\frac{dN}{dL} = N_{\star} \left (\frac{L}{L_{\star}}\right )^{\beta}  
{\rm exp}^{-(L/L_{\star})},
\end{equation}
where $N_{\star} \Gamma(1 +\beta, 1)$ is the number of galaxies with $L > L_{\star}$, and $\Gamma(-b, y)$ is the incomplete gamma function. 
This well-known functional shape provides a good analytical approximation to the measured luminosity (and/or mass) distribution of astronomical objects across a wide dynamic range. \\

We examine the shape of M83's XLFs with three methods: 
\begin{enumerate}[(i)]
\item We perform a single PL fit to the {differential} luminosity distributions, binned in intervals with an equal number of sources. This method yields the most stable and robust constraints to the power-law index of the distribution \citep{mok19}. 
\item We utilize the IDL script \texttt{MSPECFIT} \citep{mspecfit} to fit a single PL | with and without truncation | to the {un-binned, cumulative} luminosity distributions. This method is sensitive to the presence of a downturn at high luminosities, i.e., it serves to identify a characteristic luminosity above which the distribution declines sharply (if any). 
\item We perform a Maximum Likelihood (ML) fit with a Schechter function to the {un-binned} differential luminosity distributions, following \cite{mok19}. This method does not use binned data (which can hide weak features at the ends of the distribution), nor cumulative distributions (where the data points are not independent of one another). It thus gives the most robust test for the presence of a statistically significant \textit{exponential} decline at high luminosities. 
\end{enumerate}

\begin{figure*}
\centering
   \subfloat{Method (i)}{
        \includegraphics[width=\textwidth]{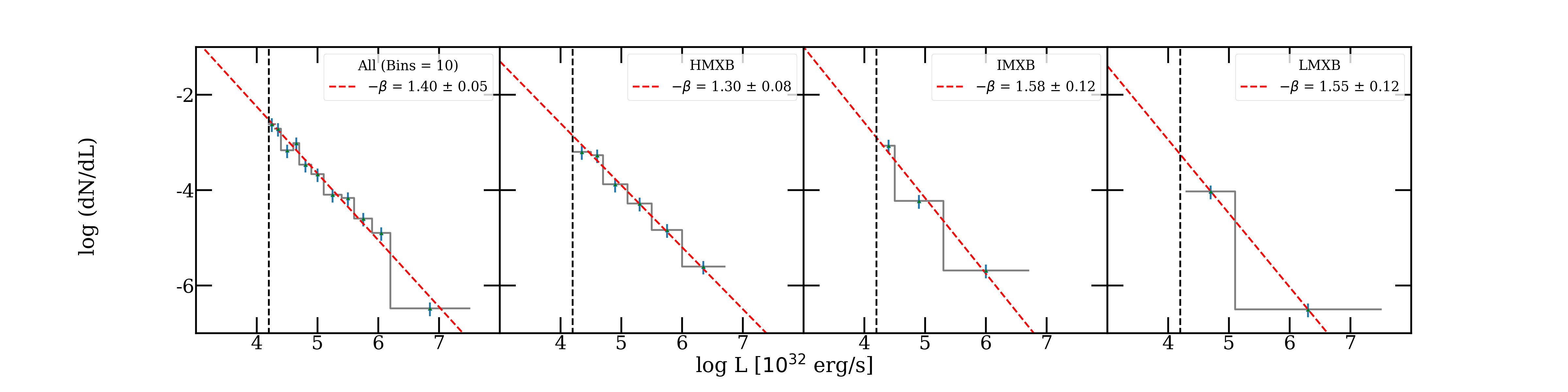}\hspace{-.55cm}
        }
    \subfloat{Method (ii)}{
         \includegraphics[width=\textwidth]{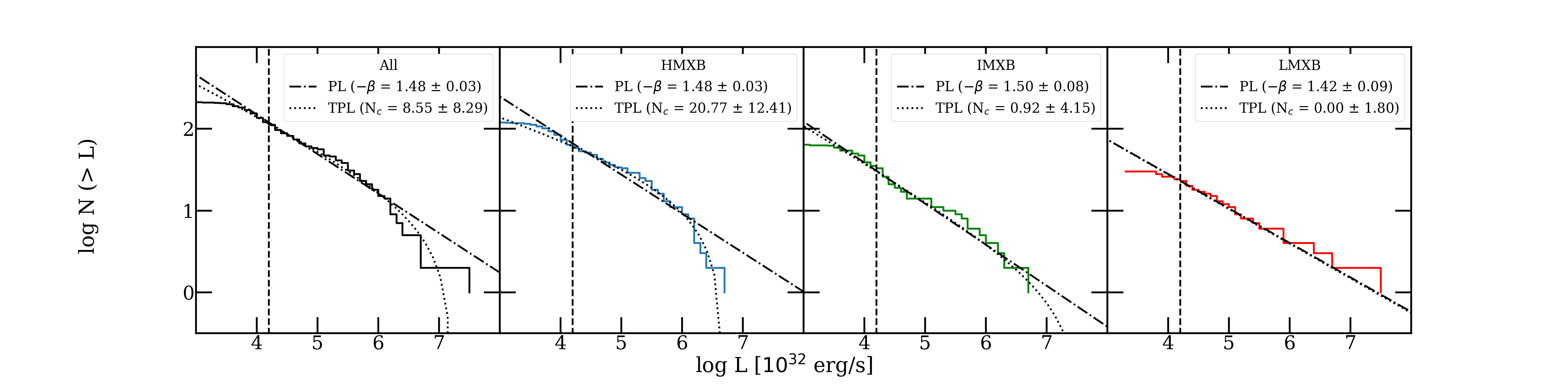}
   }
   \subfloat{Method (iii)}{
         \includegraphics[width=\textwidth]{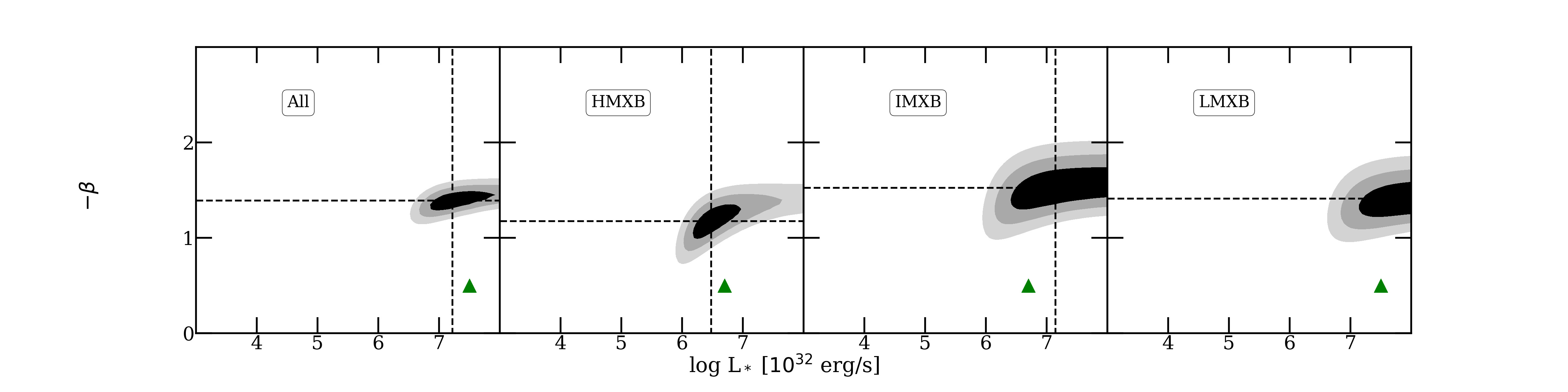}
    }
    \caption{Method (i): Fits to the differential XLFs, binned in intervals of $N=10$ sources per bin (following \citealt{mok19}). \\
    Method (ii): Fits to the cumulative, un-binned XLFs with a single power law (PL) and a truncated power law (TPL; \citealt{mspecfit}). For the top and middle panels, the dashed black vertical line indicates the 90\% completeness limit of $\ell=36.2$, above which the fits are performed.\\
    Method (iii): Maximum Likelihood fits to the cumulative XLF with a Schechter function, following \cite{mok19}. The dashed black lines indicate the best fit values for the slope and knee luminosity. Contours refer to the 1-, 2- and 3-$\sigma$ confidence levels. The green triangles indicate the luminosity of the brightest object within each sample; if the best fit knee luminosity is greater than the maximum luminosity probed by the sample, the presence of an exponential downturn is not significant.  \label{fig:XLFs}}
\end{figure*}

The top, middle and bottom panels of Figure \ref{fig:XLFs} illustrate the results of method (i), (ii), and (iii), respectively.
Unless otherwise noted, fits are performed above the 90\% completeness limit of $\ell= 36.2$ identified by L19.  

Fitting the differential XLF of M83 (composite sample) with method (i), using bins by 10 sources each, yields a slope of $- \beta=1.40\pm 0.05$. The inferred slopes for the HMXB, IMXB, and LMXB XLFs are, respectively \revise{$-\beta=1.30\pm0.08, 1.58\pm0.12$, and $1.55\pm 0.12$}, suggesting that the HMXBs are characterized by a somewhat shallower overall distribution (top panels of Figure \ref{fig:XLFs}; adopting bins of $3, 5$ and $7$ sources yields consistent results, within the uncertainties). \\

With respect to the presence of a break, method (ii) and (iii) give consistent -- and interesting -- results. Fits to the cumulative XLF with a truncated power law (TPL), with \texttt{MSPECFIT}, indicate that the \hm\ XLF is the only distribution that shows any statistically significant evidence, at the $\sim$2.5-$\sigma$ level, for a high-energy cutoff (this is indicated by values of the $N_{c}$ parameter in excess of unity in the middle panels of Figure \ref{fig:XLFs}). 

The ML fits with a Schechter function confirm this trend.
The bottom panels in Figure~\ref{fig:XLFs} show the 1-, 2-, and 3-$\sigma$ confidence contours for the best-fit values of the Schechter slope and knee luminosity. The best-fit values are indicated by the dashed black lines (the upper limit to the knee luminosity was set to be 100 times higher than that of the brightest XRB in each sample, ensuring convergence in all cases). A formally statistically significant detection of the exponential cutoff would be seen as closed 3-$\sigma$ contours in these diagrams.
We find only marginally significant evidence (at the 1- to 2-$\sigma$ level) for an exponential cutoff at the bright end of the composite XLF, suggesting that M83's XLF is formally consistent with the expectations of sampling statistics from a single power-law. This is true for the composite XLF as well as individual donor classes. 
Interestingly though, in line with the conclusions from method (ii), the presence of a (marginally significant) exponential cutoff appears to be driven by the \hm\ population, which exhibits a knee at $\ell = \revise{38.48 ^{+0.52}_{-0.33}}$ (at the 1- to 2-$\sigma$ level), whereas the \lm\ and \im\ XLFs show no evidence for a statistically significant dive (as indicated by the open 1-$\sigma$ contours in the third and fourth plots of the bottom panel of Figure~\ref{fig:XLFs}).  \\

In summary, we find that, when approximated by a single PL, the composite XRB XLF of M83 has an index of $-1.40\pm 0.05$; the shapes of the XLFs for the HMXB, LMXB, and IMXB populations show marginal deviations, with \hms\ having a shallower slope than both \lms\ and \ims. 
Our maximum likelihood fits to the Schechter function do not find formally statistically significant evidence for an exponential cutoff at the bright end of the luminosity functions. However, the presence of a high-energy cutoff in the \hm\ XLF above $\ell = 38$ (at the $\sim$2.5-$\sigma$ level) is indicated by both the ML and cumulative XLF fits.

In \S\ref{sec:xlfcomp}, we compare our fit results for the XLFs of different populations in M83 with those from previously published results for M83, as well as average XLFs derived from large samples of (star forming) galaxies.

\bigskip

\section{Discussion} \label{sec:discussion}
\subsection{The Nature of Donor Stars in IMXBs and HMXBs}\label{sec:donors}

The majority \revise{(184 out of 214)} of M83's XRBs have candidate donors that we classified as either intermediate (3-8 $M_{\odot}$) or high-mass ($>8$ \msun) stars based on the evolutionary tracks in Figure \ref{fig:CMD}. A few detected donors fall beneath the 3 \msun\ threshold and are considered \lms. 

Interestingly, we find that the majority of the intermediate- and high-mass donor stars do not follow the blue main sequence ridge that runs along the left side of the models, but rather occupy the redder portion of the evolutionary tracks. While this is likely due to a combination of effects, the majority of these objects are likely truly evolved stars.  
To start with, \hms\ are typically wind-fed (as opposed to disk-fed as a result of the donor filling its Roche lobe): the brightest of these will be those objects with more evolved donors, leading to lower surface gravity and thus higher wind loss rates. 

Indeed, the brightest, persistent HMXBs in the Milky Way (MW) and Magellanic Clouds \citep{grimm03} have either evolved or peculiar main sequence donor stars; these include, e.g., a blue supergiant in Cygnus X-1 (MW); a (high-extinction) Wolf-Rayet star in Cygnus X-3 (MW); a blue supergiant in GX 301-2 (MW); a blue supergiant in SMC X-1; a main sequence B stars with highly distorted shape in LMC X-1; and an evolved O star in LMC X-3 \citep{liu-h}. Generally speaking, about 60\% of the MW \hms\ are known or suspected Be/XRBs, while 32\% are supergiant/X-ray binaries \citep{liu-h}.
Whereas Be/XRBs in very blue bands may be expected to be close to the main sequence, their decretion disks are extremely red, and can move the overall color off of the main sequence. At the same time, since bright Be/XRBs are predominantly \textit{transients} rather than persistent, they are less likely to be represented in our investigation with respect to truly evolved stars. 

The right panel of Figure \ref{fig:CMD} shows a comparison of M83's detected donors (in black), with all of the X-ray bright\footnote{Brighter than 0.2 $\mu$Jy in the 2-10 keV range, as measured by the \textit{Rossi X-ray Timing Explorer}; this corresponds to roughly $\ell\simeq 36$ \es\ at the distance of M83.} MW XRBs listed in the \cite{liu-h} and \cite{liu-l} catalogues (blue points for \hms, red for \lms) for which measured $B$ and $V$ magnitudes are available. 
Apparent magnitudes were converted to absolute values using the formula $M = m + 5 - A_{V} - 5\rm{log} \textit{d}$, where {$d$ is the distance of the source in pc and $A_{V}$} is the interstellar extinction given by \citet{liu-h} and \citet{liu-l}. The distance to each source was approximated by the relation $A_{V} \approx 2r~\rm mag$ for sources with a galactic latitude {$b < 2^\circ$}, where $r$ is the distance in kpc.  It is reassuring that, even based on this cursory conversion, the division between \hms\ and \lms\ among MW objects aligns nicely with the expected V magnitude of \lms\ as we have defined it for M83. 
\revise{We stress that, by construction, these are intrinsically blue systems, for which extinction within the Galactic disk does not prevent a detection in the optical. As a result, the lack of Galactic systems red-ward of B-V$\simeq 0.5$ in Figure \ref{fig:CMD}  is largely a selection effect. }

\revise{At the same time, the large color spread for the M83 systems is likely compounded by intense X-ray irradiation and evolutionary effects. Specifically, a non-negligible fraction of the systems that we classify as \ims\ are likely abnormally luminous \lms.}
\revise{The initial \im\ formation rate depends on the star formation \textit{history} of the host galaxy. 
Given that M83 has been forming stars at a fairly constant rate over at least the last several 100~Myr, one would expect 
a non-negligible \textit{initial} contribution from intermediate-mass stars (and thus \ims) to the total population; in fact, the probability of forming an XRB with an initial donor between 1.5-4 \msun\ is estimated to be $\simgt 5$ times higher that of a $\simlt 1.5$ \msun\ donor \citep{pfahl03}.}
\revise{At the same time, however, population synthesis models show that for neutron star accretors (which comprise the majority of the population), intermediate-mass donors quickly evolve into low-mass stars through a short-lived thermal mass transfer phase \citep{pod02,pfahl03}. This yields a population of abnormally hot and luminous \lms, with \im\ progenitors, that may be misclassified as \ims\ through our optical method (a MW analog is the donor in the \lm\ Cygnus X-2, which has a dynamical mass of 2 \msun\ in spite of being far too luminous and hot for a low-mass sub-giant \citealt{pod99}). }

\revise{That a fraction of the donors which we identify as intermediate-mass may in fact be low-mass is also consistent with their spatial distribution (green crosses in Figure \ref{fig:sd}), as they do not seem to trace the spiral arms as well as the \hms\ (blue squares); this can be expected of systems that are longer-lived than a typical Galactic revolution time-scale of $\sim$ 250 Myr. }\\

\revise{To summarize, for M83 we conclude that (i) the published, X-ray based XLFs suffer from massive contamination from SNRs; (ii) our method confirms a non-negligible contribution from low- and possibly intermediate-mass XRBs to the total XRB XLF, i.e. between 20 and 50\%, in broad agreement with X-ray based XLFs (30\%). }


\subsection{Comparison With X-ray based XLFs}\label{sec:xlfcomp}

In this section, we compare our results first with those of L19 (\S\ref{sec:l19}) and older works (\S\ref{sec:older}).

In general, caution must be exercised when making direct comparisons with the published XLFs, for a number of reasons. First, our optical CMD-based approach enables us to \textit{directly} differentiate between different classes of XRB donors. In contrast, purely X-ray data-based XLF investigations {indirectly} differentiate between \hms\ and \lms\ by positing that the former population scales with star formation rate, and the latter with stellar mass. Furthermore, X-ray based studies do not explicitly differentiate between intermediate- vs. low- or high-mass XRB donors at all. Rather, the assumption is made that the SFR-tracing XRB population maps into truly high-mass donors ($\simgt$8 \msun). In turn, this hinges on the assumption that the adopted SFR tracer is sensitive to truly instantaneous (and by extension very short-lived) star formation episodes. Our optical reconnaissance XRB classification enables us, for the first time, to test this assumption. Since previous works have classified all XRBs into only high- or low-mass, it is not clear where the sources we classify as IMXBs end up in those studies.  In particular we compare whether the inferred number of objects in each category agree with the expectations from the published XLFs, where HMXBs are allegedly truly high-mass donors. 

Additionally, with the exception of L19, all prior X-ray based investigations of high- vs.~low-mass XRB XLFs rely on the assumption of little or no contamination to the compact X-ray population other than from cosmic background sources, which is typically minimized by limiting the search radius. While this is well justified in some cases (e.g., for massive elliptical galaxies, which are naturally devoid of \hms), it is not necessarily valid for cases such as star forming spirals, where a non-negligible fraction of the disk XRBs are likely \lms, particularly in mildly star forming galaxies. Although the contamination from X-ray emitting SNRs has also been historically neglected, \revise{independent studies \citep{long14} show that those represent a major source of contamination to the compact X-ray source population of M83 (this might apply to actively star-forming galaxies in general)}. 

\subsubsection{Comparison with L19}\label{sec:l19}

The most recent and detailed analysis of XRB XLFs to date is presented by \lem: they consider a sample of 38 nearby galaxies (including M83) spanning a vast range of morphologies and sSFRs. They use spatially resolved SFR and \mstar\ maps to divide the ($\sim$2500) \cxo-detected X-ray sources into sSFR bins and derive a global model for the scaling of the \hm\ XLF with SFR and of the \lm\ XLF with \mstar\ (accounting for the cosmic background X-ray sources with a model that scales with sky area).
In addition, they present `standard' XLF fits for each of the target galaxies; these are computed following a forward-fitting approach where the XRB and cosmic X-ray background source contributions are fit for simultaneously and convolved with a completeness function for each galaxy.
For each galaxy, the XRB contribution to the (differential) XLFs is modeled as either a single or a broken PL (see equations 4 and 5 in L19 for the adopted functional shapes). 
For practical purposes, the break and high-energy cutoff luminosity for the individual galaxy fits are fixed to $\ell_{\rm b} = {38.0}$ and $\ell_{\rm c}={40.3}$, respectively. A detailed comparison to those results, including broken power-law fits, is presented in the Appendix. Here, we focus on the broad picture, and particularly on whether there are any high-level discrepancies. \\

For M83, L19 reports a power-law index of $-1.56^{+0.05}_{-0.04}$ for the single PL fit to the composite XLF. This value is slightly steeper than the value we infer from our preferred method (i; $-\beta=1.40\pm 0.05$), as well as our method (ii; $-\beta=1.48\pm 0.03$). We suspect that the reason for this mild discrepancy has to do with the issue of SNR contamination, which we further discuss below.

Perhaps more interesting is to assess whether the XLF is best described by a single PL or exhibits any evidence for a statistically significant deviation from it (in the form or a break, downturn, or cutoff). L19 concludes that, with the exception of one target, a single PL provides a statistically acceptable fit to the data of all 38 galaxies under examination, including M83. They note that, while broken power law fits typically provide improvements to the fit statistics, in very few cases are those improvements statistically significant.
This is qualitatively consistent with our results, where the composite XLF of M83 is consistent with being sampled from a single power law, with only marginal evidence for a (\hm-driven) downturn.\\

Next, we compare the results we obtained for the \hm, \lm, and \im\ populations in M83 with the global \hm\ and \lm\ XLFs derived by \lem\footnote{For the purpose of this comparison, we refer to the best-fitting parameters from their ``Cleaned Sample" (which excludes from the sample five galaxies with low metallicity and three others with high specific number of globular clusters).}. 
Starting with the indices, the L19 \hm\ XLF has a best-fit slope of $-1.66\pm 0.02$. 
For the \lm\ XLF, the best-fit model is a broken PL with indices $-1.31^{+0.05}_{-0.07}$ and $-2.57^{+0.54}_{-0.28}$, respectively, below and above a break luminosity of $\ell_b= 38.33^{+0.21}_{-0.17}$.

Our fit to the \hm\ XLF of M83 with a single PL model, with method (i), yields a shallower index, with \revise{$\beta=-1.30\pm0.08$}. In terms of preferred functional shape, as discussed in \S\ref{sec:xlfs}, the cumulative XLF fit shows evidence (at the $\sim$ 2.5-$\sigma$ level) for a downturn in the \hm\ population, at \revise{$\ell=38.67\pm 0.18$}. This is confirmed by the ML fits with a Schechter function, which also find marginal evidence, at the 1- to 2-$\sigma$ level, for an exponential decline of the \hm\ XLF at $\ell\simeq \revise{38.48 ^{+0.52}_{-0.33}}$.

A key finding in L19 indicates that the composite \hm\ XLF has a more complex shape than previously reported; 
it exhibits a rapid decline between \lx$\simeq 10^{36}-10^{38}$ \es, a `bump' between $10^{38}-10^{40}$ \es, and an approximately exponential decline above $10^{40}$ \es. We do not see this level of complexity for the XLF of HMXBs in M83. 
Similarly, our analysis does not indicate any significant deviation from a single PL for the \lm\ XLF in M83, albeit this may again be due to low number statistics. A direct comparison with \lem\ using a broken power-law approximation of the XLF is made in the Appendix. \\

Some of the above discrepancies, such as the steeper slope obtained by \lem\ when fitting M83's total XLF, are likely driven by the high degree of SNR contamination to the M83's X-ray source population (\citealt{long14}; this is less likely to affect our conclusions regarding the presence of a downturn, since all the sources we classified as SNRs are fainter than $\ell=37.5$).  
As detailed in \S\ref{sec:nonxrb}, based on the dedicated study by \cite{long14}, we classified 103 of M83's X-ray sources as SNRs; 77 out of those 103 are brighter than the 90\% completeness limit of $\ell=36.2$, above which all fits are performed. While a detailed analysis of how this affects the measured XLF slopes for each XRB group is deferred to the Appendix, here we focus on comparing the \textit{number} of sources that we classify as \hms, \lms\ and \ims\ against the expectations from the global \hm\ and \lm\ XLFs obtained by L19\footnote{For this purpose, we adopt their best-fit values for the Cleaned sample and multiply the expected number of sources by the constant scaling factor $\omega=0.95$, inferred by L19  specifically for M83.}.

Starting with \hms, \lem\ quote a normalization value $K_{\rm HMXB}=2.06^{+0.16}_{-0.15}$ ~per~$M_{\odot}~\rm yr^{-1}$ at $\ell=38$.
By convolving the HST footprint (155,403 arcsec$^{2}$) with M83's star formation rate map (Figure \ref{fig:lehmers}b), we estimate an enclosed SFR of 2.28 $M_{\odot} \rm~yr^{-1}$. Adopting this value, M83 is then expected to have $\simeq 104$ \hms\ with $\ell \ge 36.2$. This is to be compared with the \revise{63} \hms\ identified by our optical reconnaissance analysis above the 90\% completeness limit of $\ell=36.2$. If we also consider those X-ray sources that were rejected as SNRs based on the cuts made in \S\ref{sec:nonxrb}, we obtain a total of \revise{109} objects, in good agreement with the expectations from L19. 

For \lms, we estimate that the HST footprint encloses $2.01\times 10^{10}$ \msun\ in stellar mass. With a \lm\ XLF normalization value $K_{\rm LMXB}=26.0^{+3,4}_{-2.4}$ per $10^{11}$ \msun, the L19 XLF predicts $\simeq$ 48 \lms\ above $\ell \ge 36.2$. 
We classify \revise{23} sources as \lms\ above $\ell \ge 36.2$ (or \revise{30} with the inclusion of \revise{7} X-ray sources that we rejected as SNRs). Additionally, we identify \revise{34} \ims\ above $\ell \ge 36.2$ (\revise{58} if the SNRs are included).
\revise{While the absolute numbers are less important (the global XLFs by L19 have a scatter of 0.4 dex; we include a few sources that are located outside of the SFR and stellar mass maps generated by L19; additionally, we may be too aggressive in rejecting candidate SNRs), this exercise shows that, for a galaxy with the mass and SFR of M83, based on state-of-the-art X-ray based XLF models, about 30\%\ of the detected XRBs ought to be \lms. After correcting for SNR contamination, we estimate that about 20\%\ (23/120) of the XRBs above our completeness threshold are low-mass, whereas an additional $\sim$30\%\ (34/120) are classified as intermediate-mass. However, as discussed in \S\ref{sec:donors}, a sizable fraction of this higher-mass donor population is likely made of abnormally luminous low-mass donors with intermediate mass progenitors. A quantitative assessment of the occurrence of this phenomenon in actively star-forming, nearby galaxies will be the topic of a separate paper.}

\revise{\subsubsection{Comparison With Older Works}\label{sec:older}}

\revise{In this section, we compare our results for HMXBs with those from \citet{mineo12} and \citet{sazonov17}. We defer a comparison with \cite{zhang12} for LMXBs to the Appendix, where a similar broken power-law fitting methodology as used is presented. Additional comparisons between the results detailed in L19 and those from \citet{mineo12} and \cite{zhang12} are found in L19.}

\revise{\citet{mineo12} built on the seminar works by \citet{gilfanov04} and \citet{grimm03}, which all rely on the assumption that \hm\ in star-forming galaxies can be identified by their location outside of the central bulge region, but sufficiently close that contamination by the cosmic background is not significant. They create a composite XLF for HMXBs based on sources detected in Chandra observations of 29 nearby star-forming galaxies (including M83). They make no further correction for SNR, background galaxies, or LMXBs, beyond spatial cuts. For their presumed HMXB XLF, they find a power-law index of $\beta=-1.58\pm0.02$, with a normalization $K_{\rm HMXB}=2.68\pm0.13$ per \msun~yr$\rm^{-1}$.  Their XLF predicts $\simeq 116$ \hms\ above our completeness limit and for the estimated enclosed SFR of M83 (2.28 $M_{\odot} \rm~yr^{-1}$). 
For M83, we find $\beta= -1.30 \pm 0.08$ (method i) and $\beta= -1.48 \pm 0.03$ (method ii), statistically shallower than \citet{mineo12}, with 63 HMXBs above the 90\% completeness limit. However, if we include the SNRs that were discarded from our catalog, there are 109 candidate HMXBs, which is in good agreement with the number predicted from their fits. The inclusion of SNRs also steepens our power-law fits, since SNRs tend to have fainter luminosities (see \S\ref{sec:nonxrb}).
These results indicate that at least some of the difference between our results and those found by \citet{mineo12} are driven by SNRs.}\\

\revise{More recently, \citet{sazonov17} focused on the bright end of the XLF ($L_X > 10^{38}$~erg~s$^{-1}$) for a sample of 27 nearby, star-forming galaxies. There are a number of fundamental differences in the observations and approach, which make direct comparisons challenging.  The study includes energies down to 0.25~keV, which introduces a number of super-soft sources that are unlikely to be detected in our work, that of L19, or of \citet{mineo12}. Another key difference is that they fit each X-ray spectrum to determine its unabsorbed, rather than observed, luminosity. They identify (and eliminate) a handful of foreground stars and background galaxies based on visual inspection of optical counterparts, and statistically correct for LMXB contamination by using the scaling relation from \citet{gilfanov04}. Like other works, no correction is made for SNRs.}

\revise{Despite these differences, the \citet{sazonov17} XLF has a best fit power-law index of $\beta = -1.60 \pm 0.07$, very similar to that found by L19 and \citet{mineo12}. Their scaling, however, is significantly higher | almost certainly due to the inclusion of a number of super-soft X-ray sources which are not found in the observations used to build our sample; an open question remains the possible overlap of systems that are classified as super-soft and the X-ray emitting SNRs identified by \cite{long14}.}\\

\section{Summary and Conclusions}

Building on the methodology developed by C20 for M101, we carry out an optical reconnaissance study of the XRB population in the nearby, star forming spiral galaxy M83. This method allows us to directly characterize the donor stars of each \cxo-detected compact X-ray source as low-, vs. intermediate- vs. high-mass stars \revise{(here defined as $\simlt 3$ \msun, $3-8$ \msun, and $\simgt 8$ \msun, respectively)} by comparing their donor stars to stellar evolutionary models or by estimating the ages of their parent clusters using optical photometry from multi-band high-resolution HST imaging, while also enabling a direct identification of background contaminants.  Similar to what was found by C20 for M101, we show that high-quality HST imaging of the star forming spiral M83 enables the direct detection of an optical counterpart down to about 3 \msun. 

After accounting for SNR contamination (which is especially severe in the case of M83), the differential XRB XLF of M83's (between 0.5-8 keV) is best fit by a single power law with slope $-\beta=1.40\pm 0.05$. At variance with previous studies, we also explore a Schechter function as a physically motivated alternative to the cutoff and/or broken power laws that are typically adopted to approximate XRB XLFs. Our Schechter modeling (the results of which are illustrated in the bottom panel of Figure \ref{fig:XLFs}) only identifies a marginally significant (at the 1- to 2-$\sigma$ level) exponential downturn for the \hms\ XLF in M83, $\ell\simeq \revise{38.48 ^{+0.52}_{-0.33}}$.
In contrast, the \lm\ and \im\ distributions, as well as the total XLF, are formally consistent with sampling statistics from a single power-law. 

That the \hm\ XLF in M83 deviates somewhat from a single power-law is confirmed by our cumulative distribution analysis, for which we adopt a formalism that was developed for the mass function of giant molecular clouds \citep{mspecfit}. Through this method, we identify a marginally significant truncation at \revise{$\ell= 38.67\pm 0.18$}, at the 2.5-$\sigma$ level. Again, we find that no deviations from a single power-law are required for either the \lm\ or \im\ population. \\

Last, our optical reconnaissance methodology enables us, for the first time, to make direct inferences on the role of \ims\ in the XRB XLFs. 
The assumption that the SFR-tracing XRB population maps into truly high-mass donors is typically predicated upon the notion that the adopted SFR tracer is sensitive to instantaneous, and hence very-short lived, star formation episodes. However, we note that the survival rate of \ims\ is arguably dependent on the host galaxy star formation \textit{history}. For spiral galaxies like M83, which had fairly constant (high) rates of star formation over at least the last Gyr \citep{chandar10}, \ims\ are likely to yield a non-negligible contribution to the XRB population. 
\revise{At the same time, X-ray binary evolutionary models show that, after sustaining a highly super-Eddington mass loss phase on thermal time scales, these will quickly evolve into low-mass donors with unusually hot spectral types. Owing to this effect, we estimate a non-negligible contribution from low- and possibly intermediate-mass XRBs to the global XLF of the star-forming galaxy M83, i.e. between 20 and 50\% (to be compared with an estimated contribution from \lms\ at the 30 per cent level based on the L19 X-ray XLFs).}
\revise{Finally, we caution against a sizable contribution from X-ray emitting SNRs to the published XLFs for M83, and possibly other star-forming galaxies (for the case of M83, more than 30\% of the compact X-ray sources that fall within the HST footprint have been identified by \citealt{long14} as SNRs).}

In future papers, we will extend our methodology to a sample of several tens of nearby galaxies with high-quality HST and \cxo\ coverage, so as to increase our number statistics and deliver a direct census of the XRB population based on our novel optical reconnaissance of the donor type. 

\acknowledgements
QH is partially funded by a Rackham Merit Fellowship, awarded by the University of Michigan Rackham Graduate School. We are grateful to Tom Maccarone for useful comments and suggestions.

\appendix
\setcounter{figure}{0} \renewcommand{\thefigure}{A.\arabic{figure}}
\setcounter{table}{0} \renewcommand{\thetable}{A.\arabic{table}}
%

\section{Single vs. Broken Power-law fits}

\begin{figure*}
\centering
\includegraphics[width=0.335\textwidth]{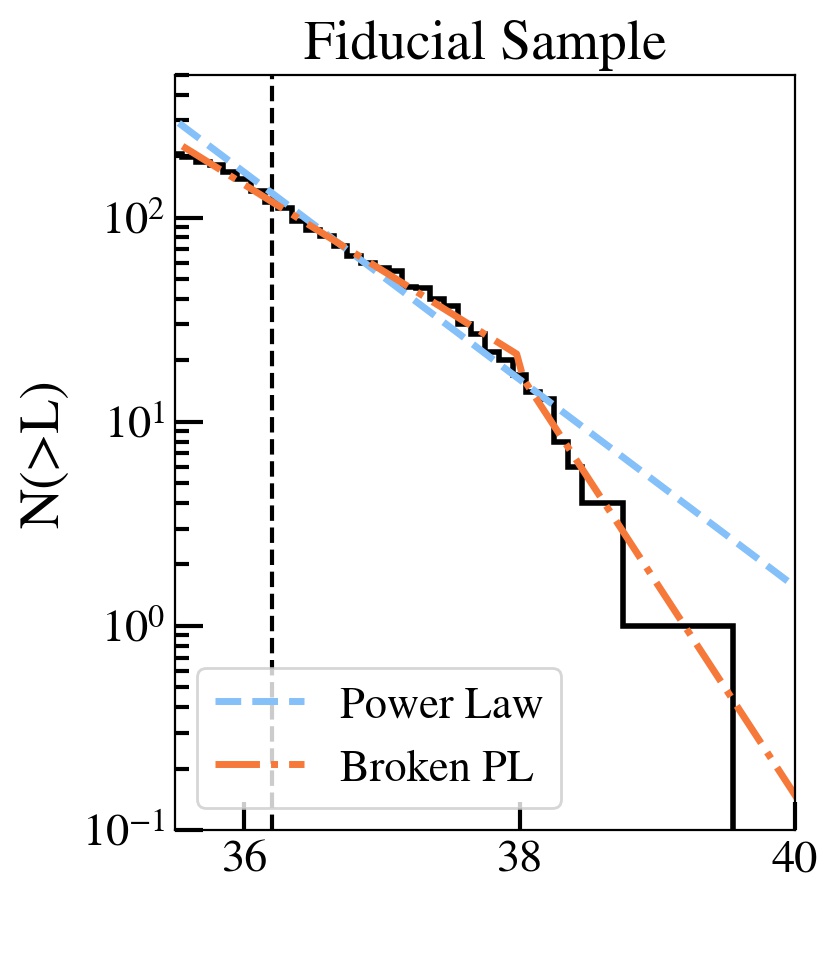}
\includegraphics[width=0.3\textwidth]{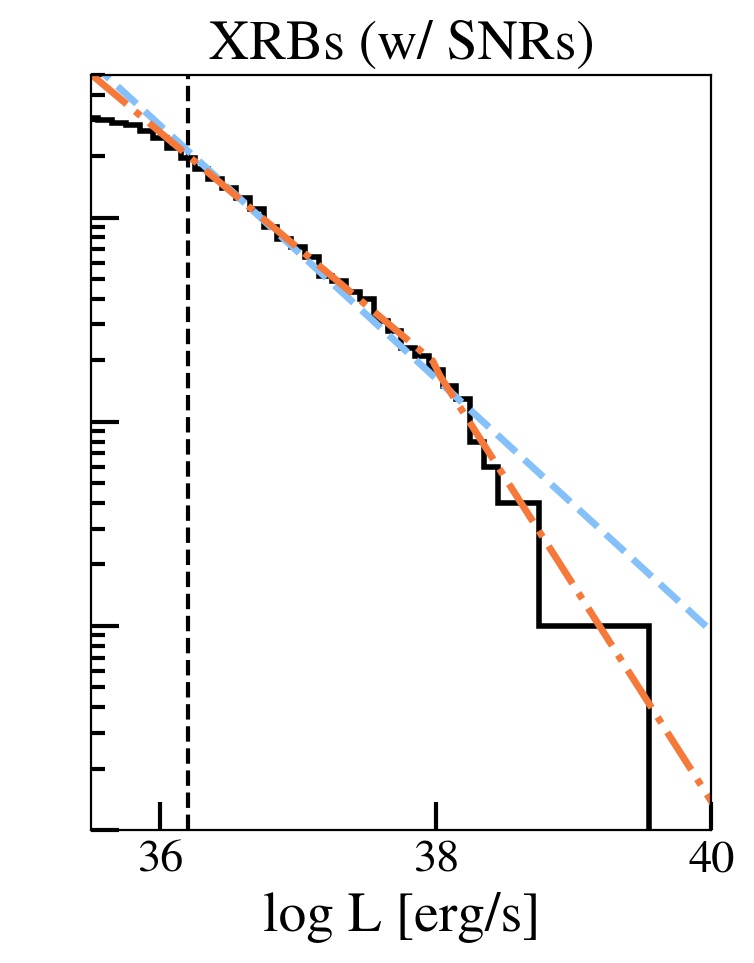}
\includegraphics[width=0.3\textwidth]{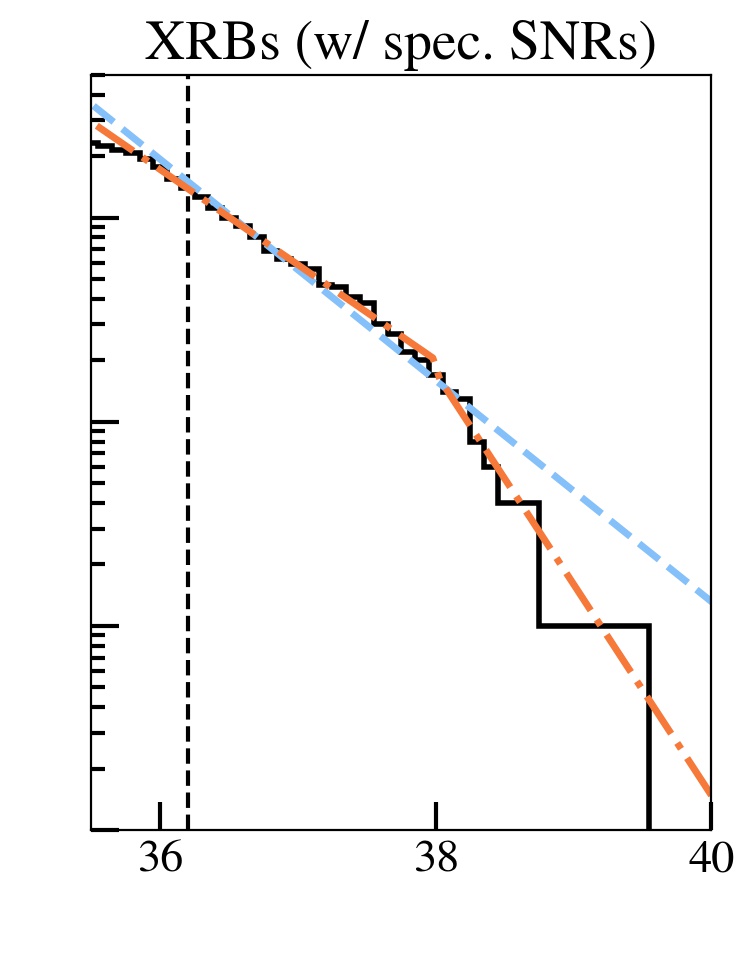}
\caption{Single and broken power-law function fits to the composite XRB XLFs in M83 (i.e., the ``All XRBs" rows) of the fiducial sample containing all XRBs with luminosities above the 90\% completeness limit (left), sources including XRBs and all SNR (center), and sources with only SNRs identified in \citet{long14} removed (right). See Table \ref{tab:XLF_unbinned} for the corresponding best-fit parameters. \label{fig:XLFcumul} }
\end{figure*}

To facilitate a direct comparison with the literature, where the XRB XLFs are typically fit with single or broken power-laws (PLs and BPLs), here we approximate M83's XRB luminosity distribution as follows:

\begin{equation}\label{eq:pl}
    N(>L)_{\rm PL} = \revise{\kappa}_{\rm PL}L^{-(\alpha - 1)}
\end{equation}

\begin{equation}\label{eq:bpl}
    N(>L)_{\rm BPL} = \left\{\begin{array}{ll}{\revise{\kappa}_{\rm BP1}L^{-(\alpha_{1} - 1)},} & {(L \leq L_{\rm b})} \\ {\revise{\kappa}_{\rm BP2}L^{-(\alpha_{2} - 1)},} & {(L > L_{\rm b})}\end{array}\right.
\end{equation}

\noindent where $N(>L)$ is the cumulative XLF, $L_{\rm b}$ is the break luminosity of the broken power-law (BPL), and the \revise{$\kappa$} values are normalization constants \revise{(note, $\kappa$ is defined differently than the normalization constants of the differential XLF fits, K, discussed in \S\ref{sec:discussion}). All luminosities are in units log \es.} We choose to fit un-binned cumulative distributions here, as they tend to be more sensitive to the presence of a downturn at high luminosities. At the same time, the choice to represent the power-law indices as ($\alpha - 1$) is meant to facilitate a direct comparison to those studies that adopt a differential, rather than cumulative, form of the XLF. Apart from the break luminosity, which is fixed to $\ell_b = 38.0$ for consistency with L19, all variables are fitted for, using the Python \texttt{scipy.curve\_fit} function. We are primarily interested in comparing our fits to the results reported by L19. However, since previous studies do not take into account the SNR contamination to the compact X-ray source population, we also examine the effects of removing (secure and candidate) SNRs to the shape of the XLF. For completeness, we also report the results of fitting the \hm, \lm, and \im\ samples separately.
Table \ref{tab:XLF_unbinned} and Figure \ref{fig:XLFcumul} summarize the results of our fits. 

Our ``Fiducial Sample" (from the main Paper, sample a) includes a total of \revise{120} sources above the 90\% completeness limit of $\ell=36.2$. This excludes 103 sources identified as SNRs either by \citet[][76 in total]{long14} or by their X-ray properties (27 additional sources), as described in \S\ref{sec:nonxrb}. For this sample, we find a PL index of $1.51 \pm 0.02$, and BPL indices of $1.42 \pm 0.01$ and $2.03 \pm 0.07$, respectively below and above the break. This is to be compared with the values inferred by \lem: a PL index of $1.56^{+0.05}_{-0.04}$, and BPL indices of $1.47^{+0.06}_{-0.05}$ and $1.93^{+0.22}_{-0.18}$ below and above the break, all of which are formally consistent with our best-fit values, within the uncertainties.  

Including the SNRs to the sample (``With SNRs", sample b) yields a single PL index \revise{$1.62 \pm 0.02$}, whereas the BPL indices are $1.57 \pm 0.01$ and $2.06 \pm 0.07$.  Not surprisingly, the inclusion of these (low-luminosity) sources slightly steepens the inferred PL slope, as well as the BPL slope below the break. 
The X-ray based diagnostics we developed to identify and reject SNR candidates (see \S\ref{sec:nonxrb}) may be too aggressive in that it may lead to the rejection of faint, X-ray soft XRBs. 
To fully illustrate the effects of our rejection criteria, we present the results of our fits to a sample in which \textit{only} the 76 X-ray sources that were directly identified by \citet{long14} are removed (``With spec. SNRs", sample c), while the additional 27 SNR candidates that we reject from the fiducial sample on the basis of the X-ray color are \emph{included} in the XLF. 
However, it should be noted that, in terms of the inferred slopes, all three fits are consistent with the values reported by \lem, within 2-$\sigma$. 
This suggests that, though strict, our SNR filtering method does not drastically alter the overall shape of the XLF. It does, however, alter the normalization. 

\revise{For completeness, we compare our BPL fits to \citet{zhang12}, which studies the \lm\ XLF using a sample of 20 nearby elliptical galaxies. A comparison to this study is also conducted in L19. They fit the XLF to a BPL model with two breaks (at $5.5 \times 10^{37}$ and $6\times10^{38}$ \es), as opposed to a single break used both by L19. 
L19 find no improvement in the quality of the fit when adopting two breaks; therefore, they focus only on the parameters around the first break, a prescription we follow here. The indices of the \lm\ XLF found by \citet{zhang12} are $1.02^{+0.07}_{-0.08}$ and $2.06^{+0.06}_{-0.05}$, which yields $\sim32$ expected \lms\ within the HST footprint of M83. By our methods, we find 23 \lms\ in our fiducial sample (a), 30 \lms\ in sample (b), and 26 \lms\ in sample (c). Like other studies that indirectly estimate the contribution of \lms\ in late-type galaxies, the \citet{zhang12} XLF most likely includes SNR contamination, as well as contamination from \ims. Like L19, we obtain slopes that are steeper in the faint end by a statistically significant margin for all fits. At higher luminosities, however, rather than steepening, our \lm\ XLF becomes much shallower. There are two possible explanations for these observations: first, this may be due to the fact that M83 is a late-type galaxy, and at higher sSFRs, young \lms\ may achieve higher luminosities (\citealt{fragos13, kim10, lehmer14, lehmer17}; L19); or second, a BPL is simply a poor representation of the \lm\ XLF, as demonstrated in \S\ref{sec:xlfs}}.

\begin{deluxetable*}{rccccccccc}[b]
\tabletypesize{\footnotesize}
\tablewidth{0pt}
\tablecaption{XRB XLF fit parameters\label{tab:XLF_unbinned}}

\def \lum {$\rm erg~s^{-1}$}
\tablehead{
 \colhead{} & \multicolumn{2}{c}{\sc Power Law} & \multicolumn{4}{c}{\sc Broken Power Law$^\dagger$}   \\
\vspace{-0.3in} \\
\colhead{} & \multicolumn{2}{c}{\rule{1.5in}{0.01in}} & \multicolumn{4}{c}{\rule{3.0in}{0.01in}} \\
\vspace{-0.25in} \\
\multicolumn{1}{c}{\sc XLF} & \colhead{$\kappa_{\rm PL}$} & \colhead{$\alpha$} & \colhead{$\kappa_{\rm BP1}$} & \colhead{$\kappa_{\rm BP2}$} & \colhead{$\alpha_1$} & \colhead{$\alpha_2$}   \\
\vspace{-0.25in} \\
\multicolumn{1}{c}{(1)} & \colhead{(2)} & \colhead{(3)} & \colhead{(4)} & \colhead{(5)} & \colhead{(6)} & \colhead{(7)} \\
\vspace{-0.23in} 
}
\startdata
\textit{(a) Fiducial Sample}\\
       All XRBs &   60.89 $\pm$ 0.86    &       1.51 $\pm$ 0.02     &   58.43 $\pm$ 0.41    &   79.74 $\pm$ 2.70    &   1.42 $\pm$ 0.01     &   2.03 $\pm$ 0.07      \\ 
       HMXBs    &   58.71 $\pm$ 1.12    &       1.46 $\pm$ 0.03     &   56.82 $\pm$ 0.69    &   102.89 $\pm$ 7.34    &   1.39 $\pm$ 0.02     &   2.66 $\pm$ 0.19      \\ 
       IMXBs    &   59.22 $\pm$ 0.85    &       1.51 $\pm$ 0.02     &   57.79 $\pm$ 0.91    &   77.90 $\pm$ 5.12   &   1.47 $\pm$ 0.02     &   2.02 $\pm$ 0.13      \\ 
       LMXBs    &   56.97 $\pm$ 0.32    &       1.45 $\pm$ 0.01     &   56.57 $\pm$ 0.46    &   56.19 $\pm$ 1.80   &    1.44 $\pm$ 0.01     &   1.43 $\pm$ 0.05     \smallskip\\
\textit{(b) With SNRs}\\
       All XRBs &   64.71 $\pm$ 0.61    &       1.62 $\pm$ 0.02     &   63.07 $\pm$ 0.28    &   80.67 $\pm$ 2.70    &   1.57 $\pm$ 0.01     &   2.06 $\pm$ 0.07   \\
       HMXBs    &   62.92 $\pm$ 0.86    &       1.58 $\pm$ 0.02     &   61.57 $\pm$ 0.53    &   106.34 $\pm$ 7.01    &   1.54 $\pm$ 0.01     &   2.75 $\pm$ 0.18   \\
\enddata
\tablecomments{The best-fit parameters inferred by approximating the XLFs with the functional shapes given in Equations \ref{eq:pl} and \ref{eq:bpl}.
$^\dagger$For the broken power-law fits, the break luminosity is fixed at $\ell_b = 38$, to facilitate comparison with \citet{lehmer19}. Three subsamples are examined: (a) the fiducial sample  of XRBs identified in the main Paper; (b) the fiducial sample plus \textit{all} SNRs; and (c) the fiducial sample plus \textit{only} SNRs identified in \S\ref{sec:nonxrb} using our L$_{X}$-HR criterion, with all SNRs identified in \citet{long14} removed. \\
Columns (1)-(10) describe: \\
(1) the population of XRBs fit by each function; (2) the normalization of the power-law; (3) the index of the power-law; \\
(4) the normalization of the power-law fit to the XLF below $\ell_b$; (5) the normalization of the power-law fit to the XLF above $\ell_b$;\\
(6) the index of the power-law fit to the XLF below $\ell_b$; (7) the index of the power-law fit to the XLF above $\ell_b$.}
\end{deluxetable*}

\clearpage
\startlongtable
\begin{deluxetable*}{cccccccccccc}
\tabletypesize{\footnotesize}
\tablewidth{0pt}
\tablecaption{Properties and Classifications of M83 X-ray Sources \label{tab:allsources}}
\tablehead{     
\colhead{ID} & \colhead{CSC ID} & \colhead{Long ID} &  \colhead{R.A.} &  \colhead{Dec} & \colhead{\lx} &          \colhead{V} &    \colhead{B-V} &   \colhead{V-I}& \colhead{Class} &  \colhead{Cluster Age} & \colhead{CF}}
\startdata
 L19X048 &  2CXO J133648.3-295244 &    X046 &  204.201273 &   -29.879037 &               36.4 &  -7.180338 & -0.119078 &  0.196804 &            SNR &           | &  2.0 \\
 L19X049 &  2CXO J133648.2-295136 &    X047 &  204.201358 &   -29.860496 &               36.0 &  -2.488338 &  2.202922 &  1.067804 &           IMXB &           | &  2.0 \\
 L19X050 &  2CXO J133648.7-295229 &    X048 &  204.203291 &   -29.874848 &               37.1 &  -6.195338 & -0.051078 &  0.157804 &           HMXB &           | &  1.0 \\
 L19X051 &  2CXO J133649.1-295258 &    X049 &  204.204673 &   -29.882838 &               37.6 &  -5.178338 &  0.190922 &  0.666804 &           HMXB &           | &  3.0 \\
 L19X052 &  2CXO J133649.1-295125 &    X050 &  204.204964 &   -29.856996 &               36.3 &        | &       | &       | &           LMXB &           | &  1.0 \\
 L19X053 &  2CXO J133649.2-295303 &    X051 &  204.205104 &   -29.884200 &               37.5 &  -3.026338 & -0.191078 &  0.429804 &   \textit{SNR} &           | &  2.0 \\
 L19X055 &  2CXO J133649.4-295014 &    X052 &  204.205935 &   -29.837387 &               36.7 &  -3.747338 &  0.325922 &  0.184804 &           IMXB &           | &  1.0 \\
 L19X056 &  2CXO J133649.7-295217 &    X053 &  204.207525 &   -29.871477 &               37.0 &  -3.664338 & -0.396078 & -0.012196 &            SNR &           | &  1.0 \\
\enddata
\tablecomments{Properties and classifications of all M83 X-ray sources identified in \citet{lehmer19} that fall within the footprint of the HST image. For each source, the ID indicates the ID number assigned in \citet{lehmer19}, while CSC ID is the full ID from the \emph{Chandra} Source Catalog Release 2, and Long ID is the ID from \citet{long14}, where applicable. X-ray Luminosities are in units log \es, and magnitudes are absolute mags estimated at a distance of 4.61 Mpc. The classification for each source is given, with italics representing SNRs identified using our HR-L$_{\rm{X}}$ criterion or XRBs associated with clusters. Classifications in parentheses are objects with uncertain ``candidate" classifications, as reported in \citet{long14} or as found by our methods. Sources that were identified as candidate quasars are labeled `Gal.' For XRBs in clusters, the cluster ages are given in units log years. A confidence flag (CF) is assigned to each source based on the ``strength" of the identification of the X-ray emitter: a CF of 1 represents the most certain classifications (those determined in other studies, or XRBs with a clear donor, with multiple candidates of similar mass, or a clear absence of a donor); CF ratings of 2 or 3 may indicate that a source is in a dust-obscured region, such as near the nucleus or along a dust lane (since the presence of heavy dust could potentially mask high-mass stars, background galaxies, and clusters, leading to possible mis-identifications), or that there are multiple sources of different masses within the 2-$\sigma$ radius.} 
\end{deluxetable*}

\begin{figure*}
\includegraphics[width=\linewidth]{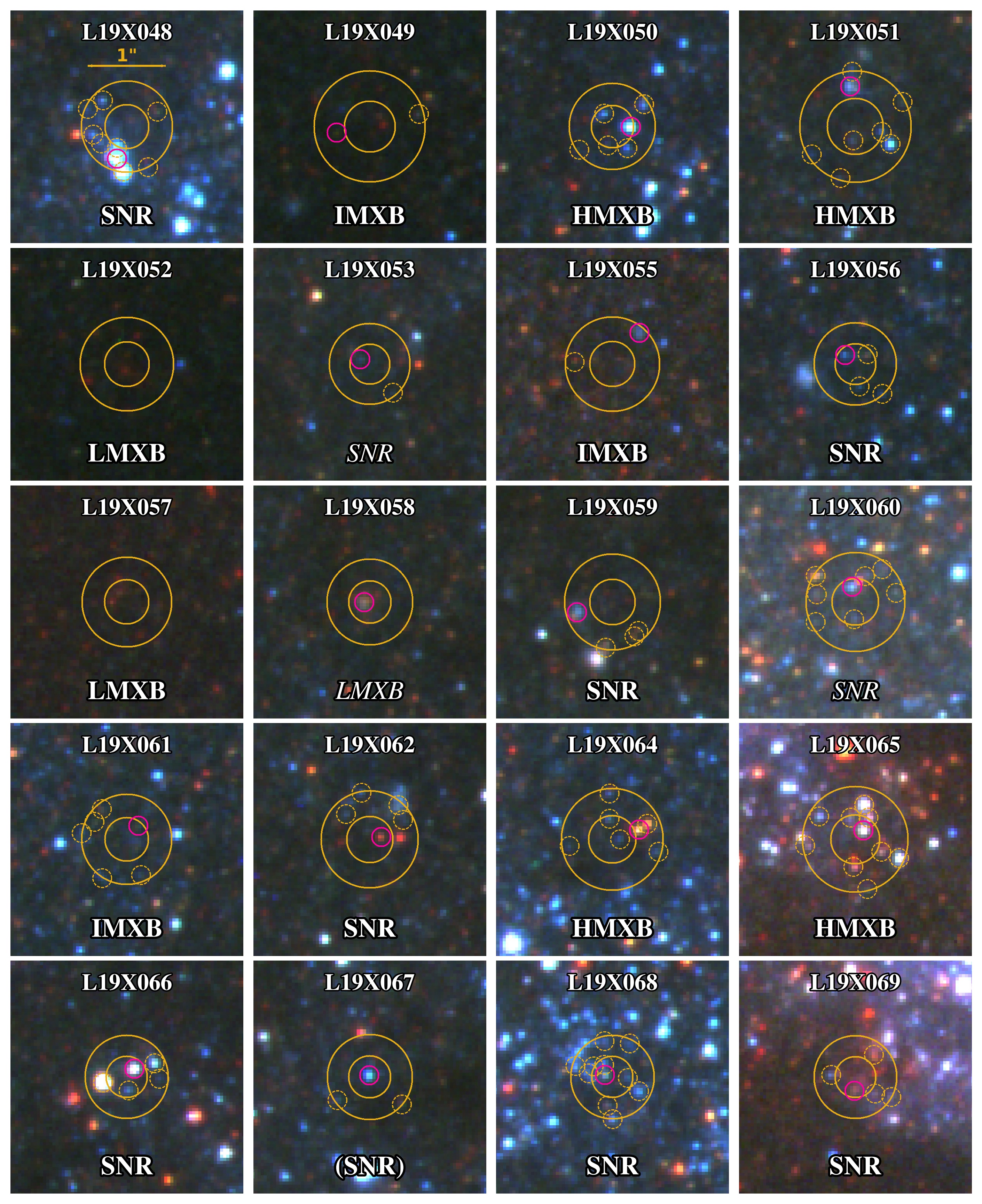}
\caption{Identified point sources within the 1- and 2-$\sigma$ radii for each X-ray source, with the most likely donor circled in red. Classifications in italics represent SNRs identified using our HR-L$_{\rm{X}}$ criterion or XRBs associated with clusters. Parentheses indicate objects with uncertain ``candidate" classifications, as reported in \citet{long14} or as found by our methods. Sources that were identified as candidate quasars are labeled `Gal.' Note: All images are set to scale with source L19X048, except L19X206, for which a unique scale is shown.\label{fig:mosaics}}
\end{figure*}

\begin{figure*}
\ContinuedFloat
\includegraphics[width=\linewidth]{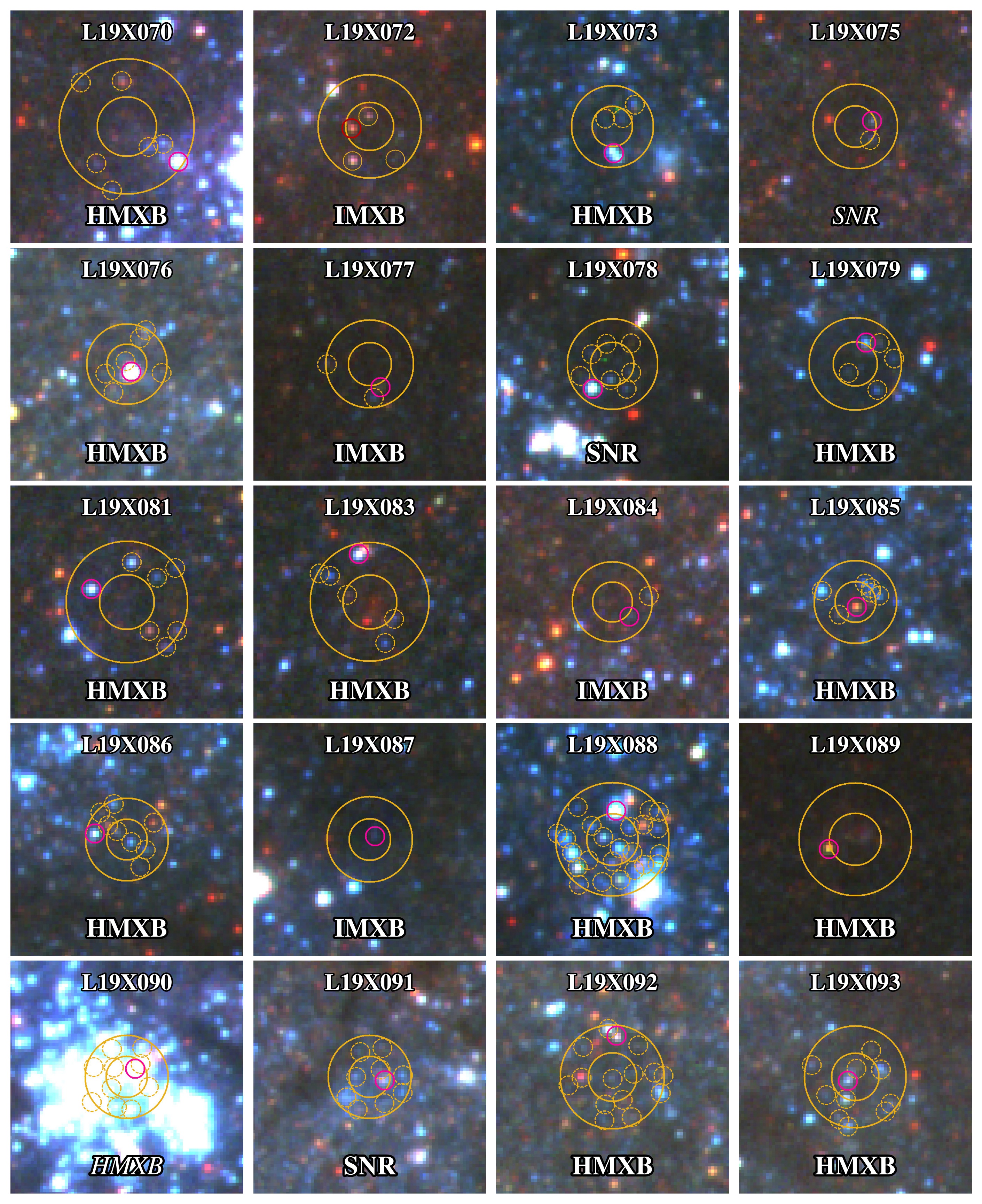}
\caption{\textit{(continued)}}
\end{figure*}

  \begin{figure*}
\ContinuedFloat
\includegraphics[width=\linewidth]{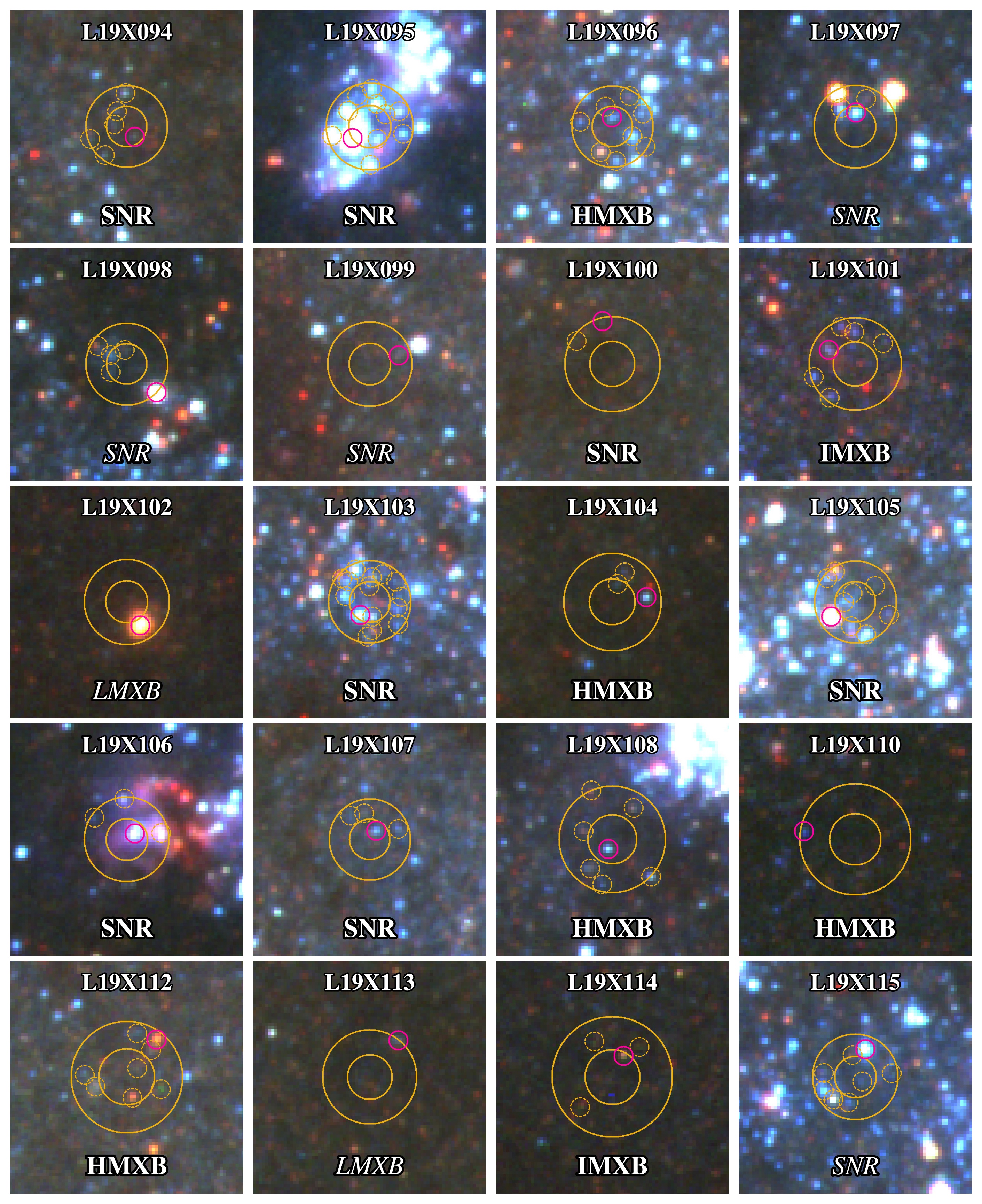}
\caption{\textit{(continued)}}
\end{figure*}

  \begin{figure*}
\ContinuedFloat
\includegraphics[width=\linewidth]{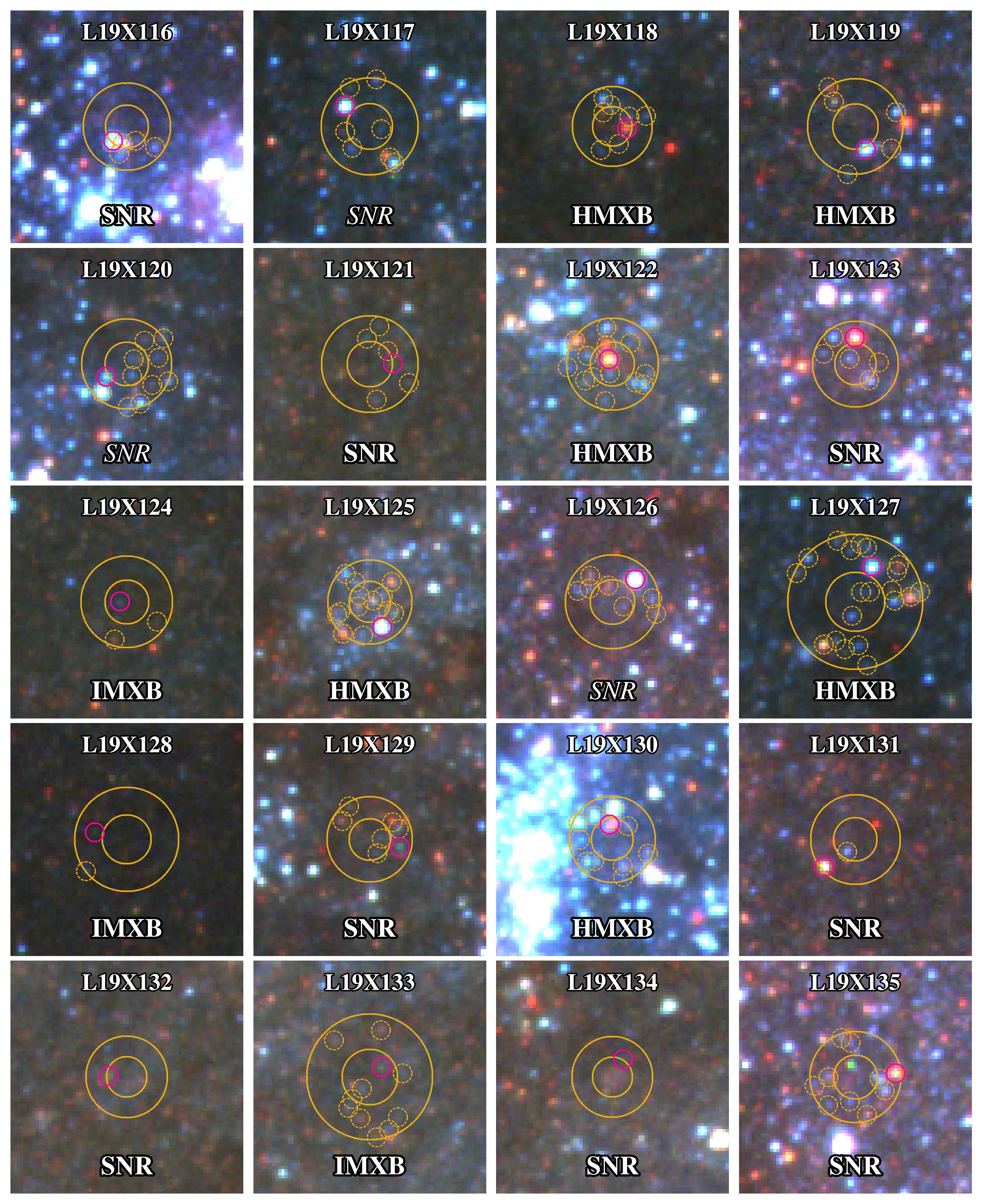}
\caption{\textit{(continued)}}
\end{figure*}

   \begin{figure*}
 \ContinuedFloat
\includegraphics[width=\linewidth]{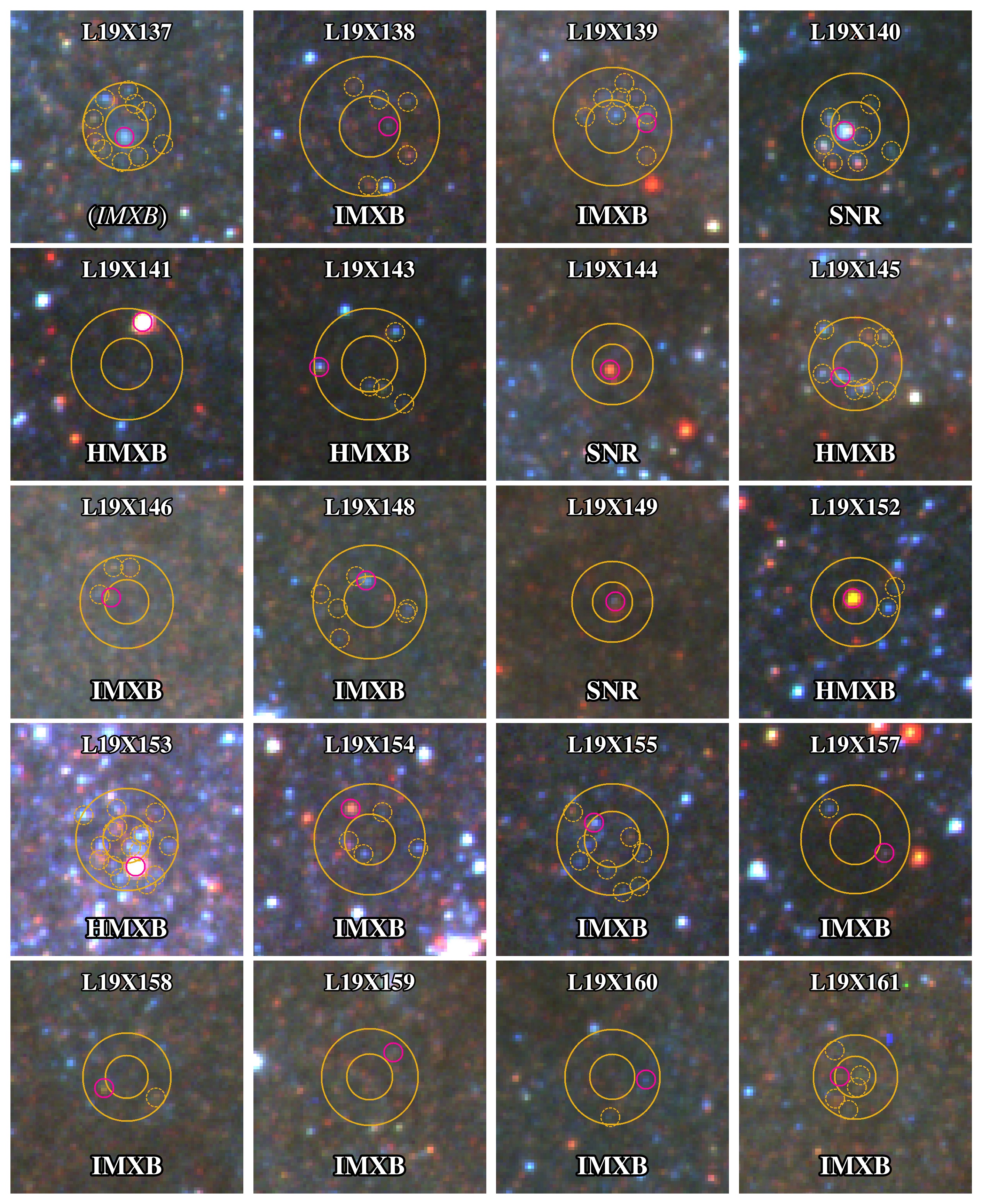}
\caption{\textit{(continued)}}
\end{figure*} 

  \begin{figure*}
\ContinuedFloat
\includegraphics[width=\linewidth]{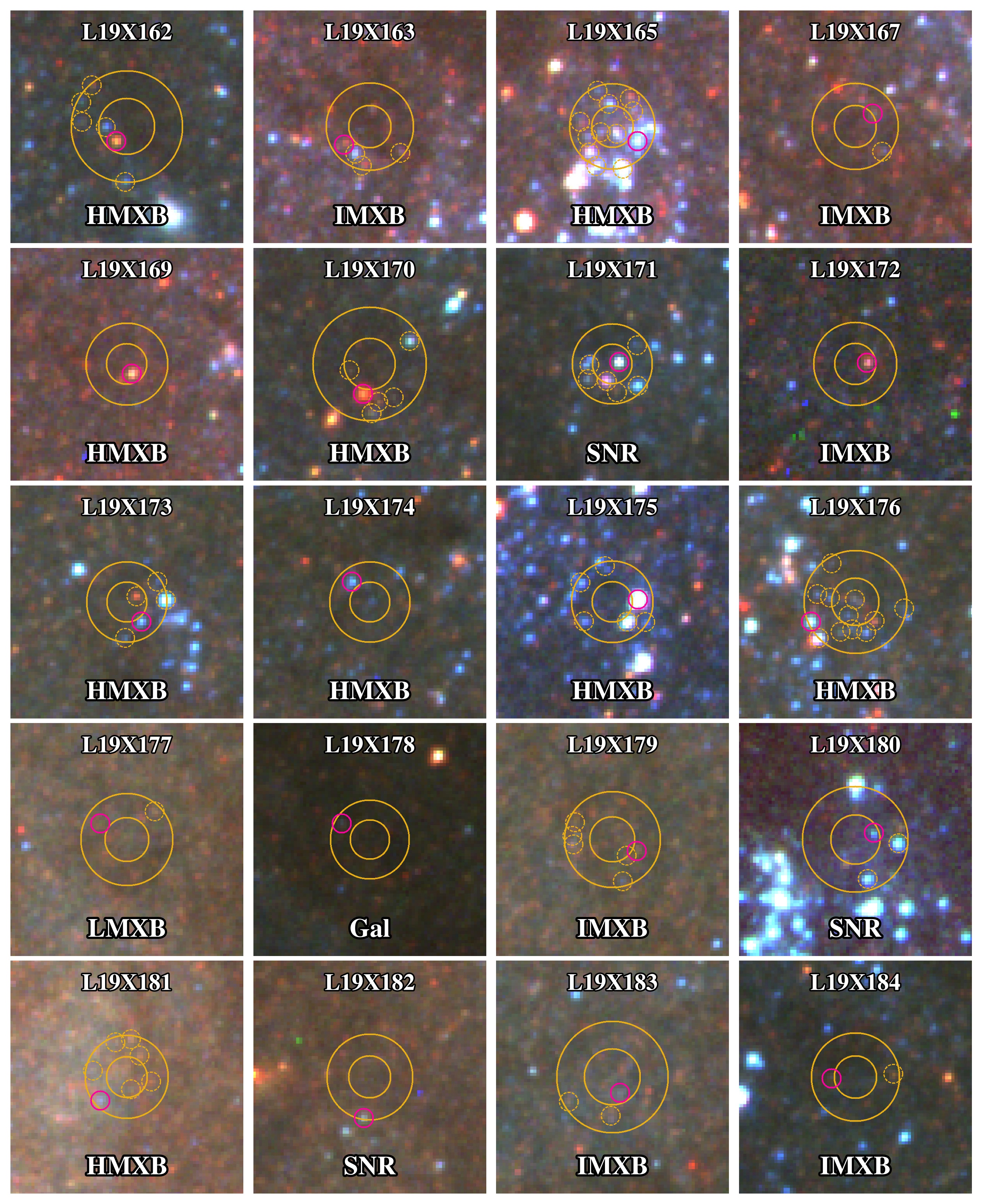}
\caption{\textit{(continued)}}
\end{figure*}   

  \begin{figure*}
\ContinuedFloat
\includegraphics[width=\linewidth]{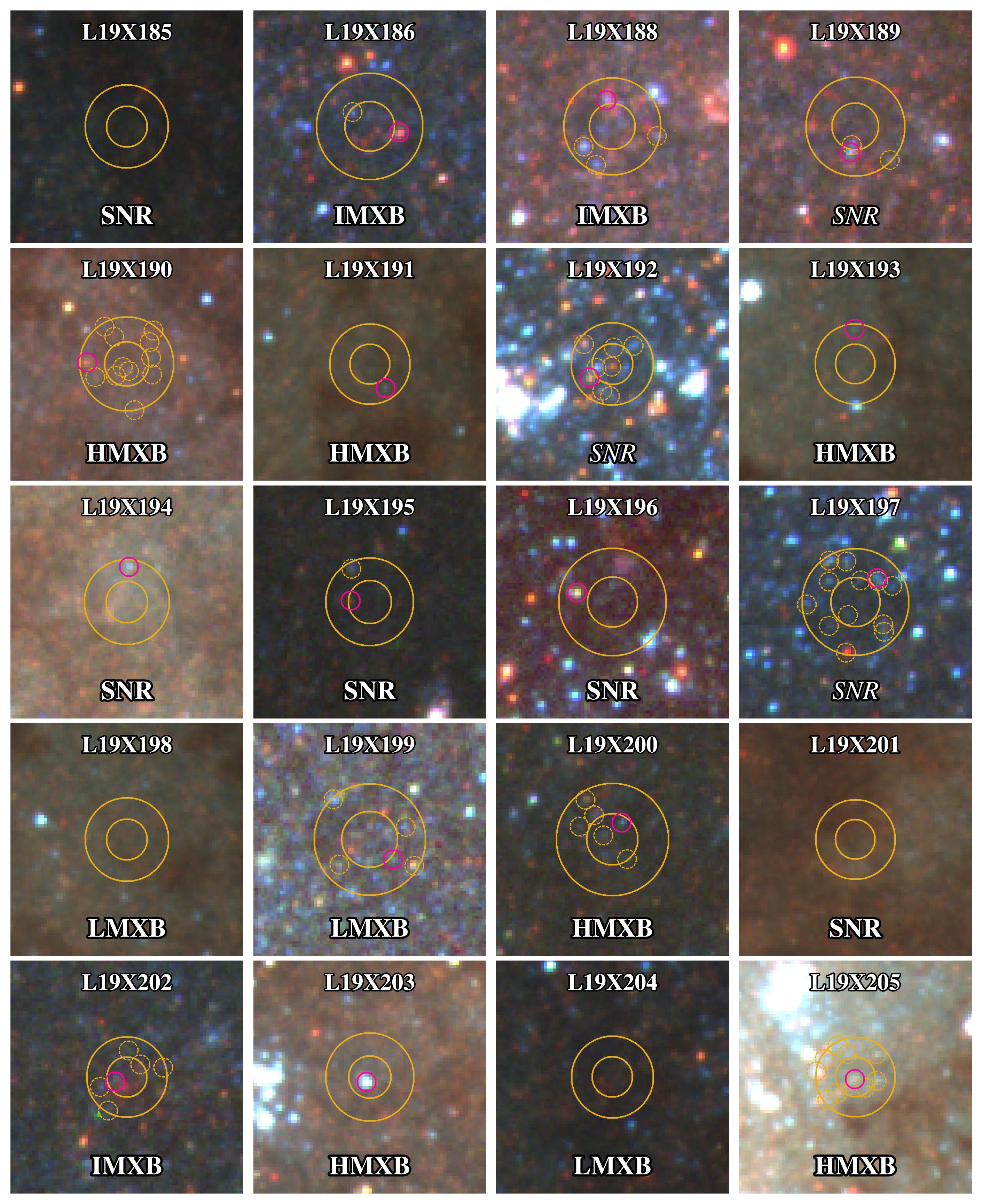}
\caption{\textit{(continued)}}
\end{figure*}

  \begin{figure*}
\ContinuedFloat
\includegraphics[width=\linewidth]{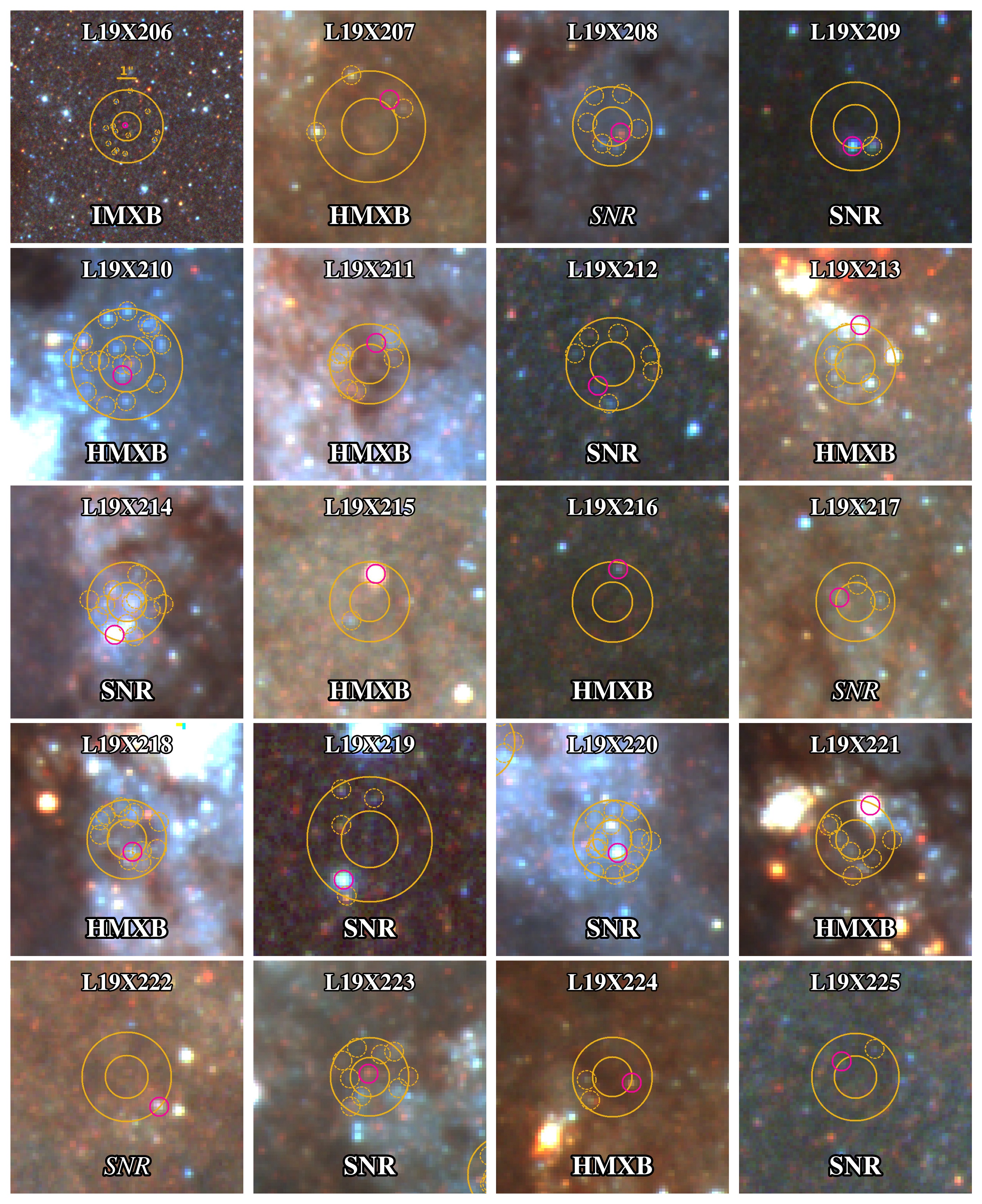}
\caption{\textit{(continued)}}
\end{figure*}

   \begin{figure*}
 \ContinuedFloat
\includegraphics[width=\linewidth]{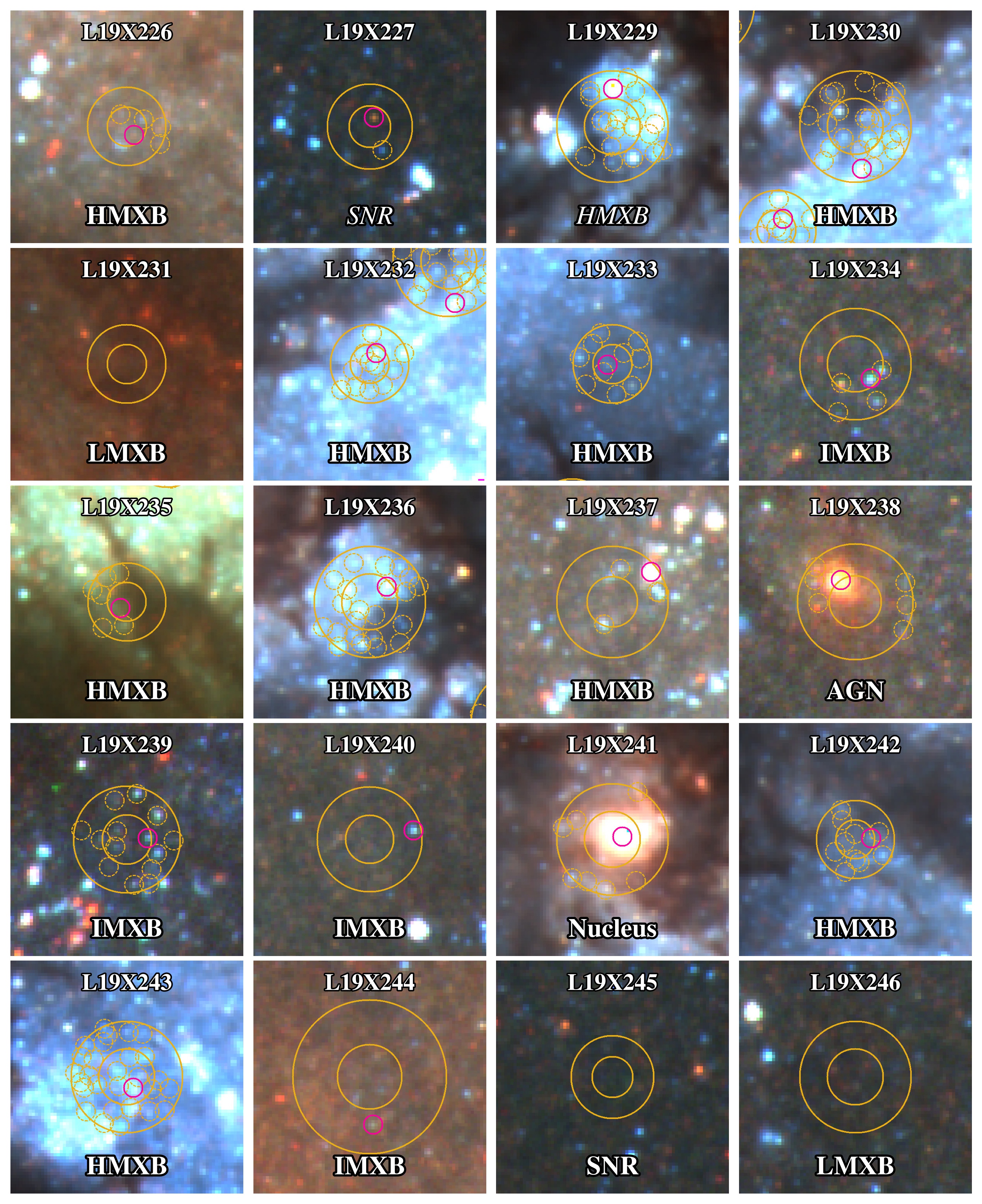}
\caption{\textit{(continued)}}
\end{figure*}   

  \begin{figure*}
\ContinuedFloat
\includegraphics[width=\linewidth]{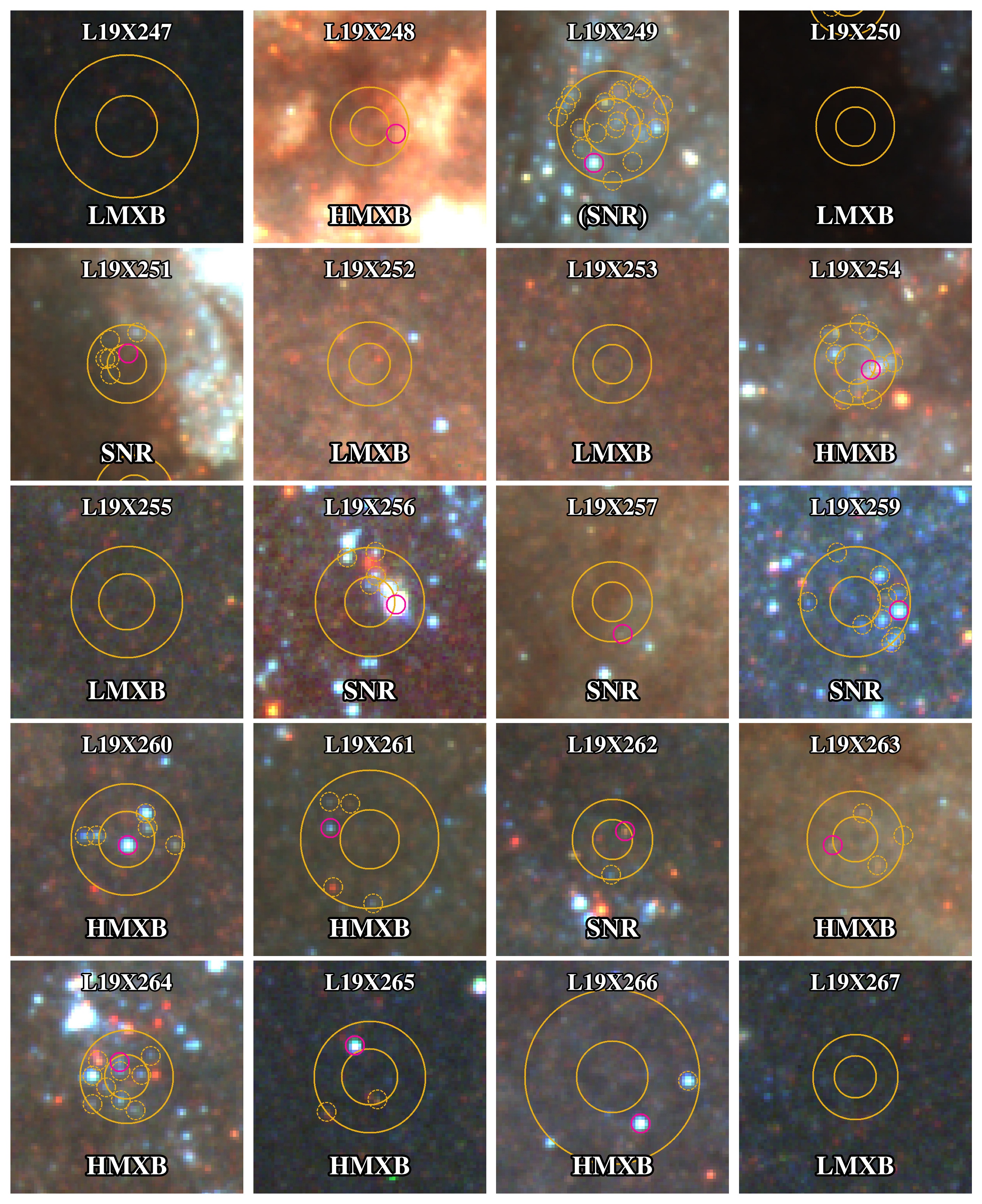}
\caption{\textit{(continued)}}
\end{figure*}

  \begin{figure*}
\ContinuedFloat
\includegraphics[width=\linewidth]{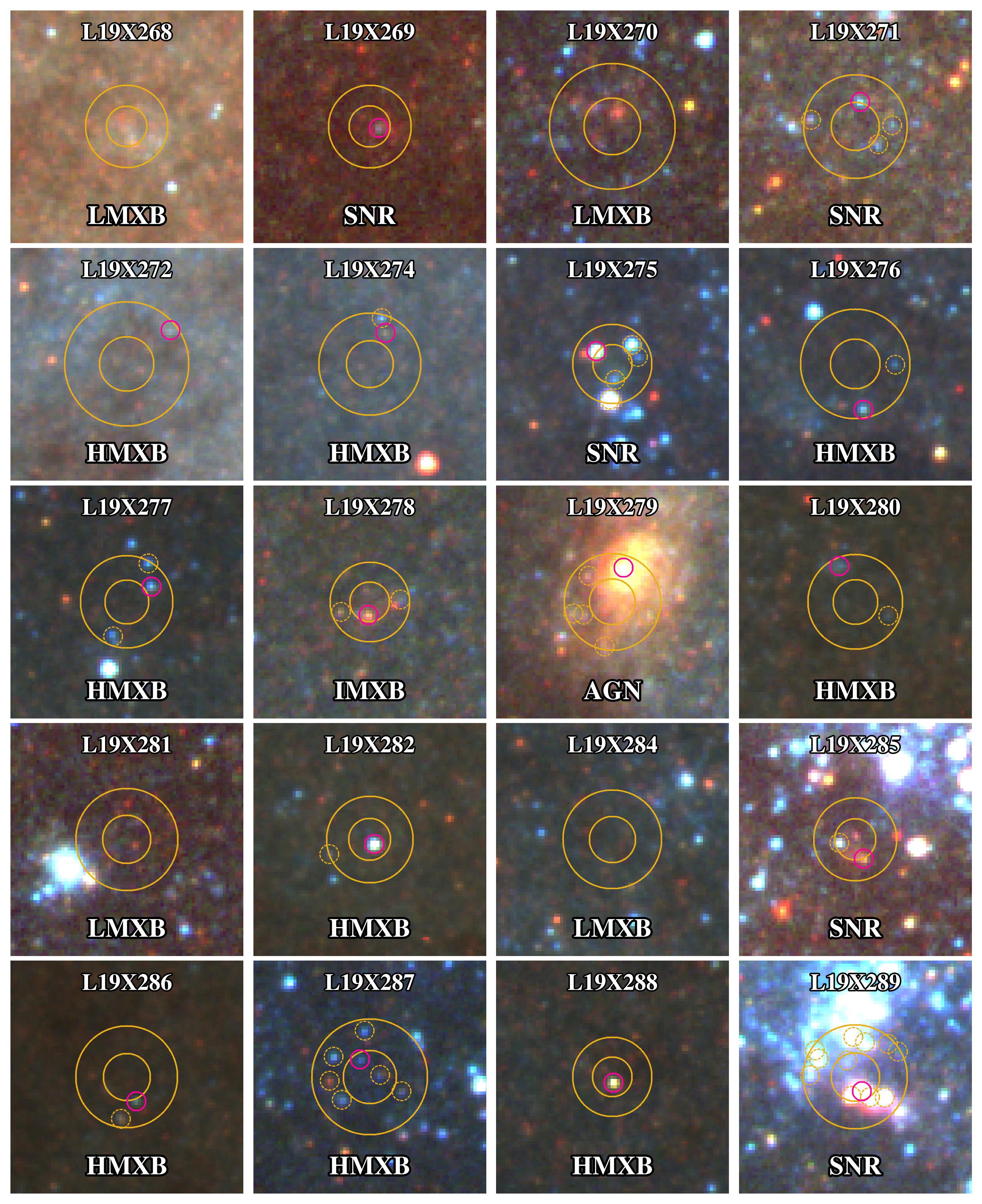}
\caption{\textit{(continued)}}
\end{figure*}

  \begin{figure*}
\ContinuedFloat
\includegraphics[width=\linewidth]{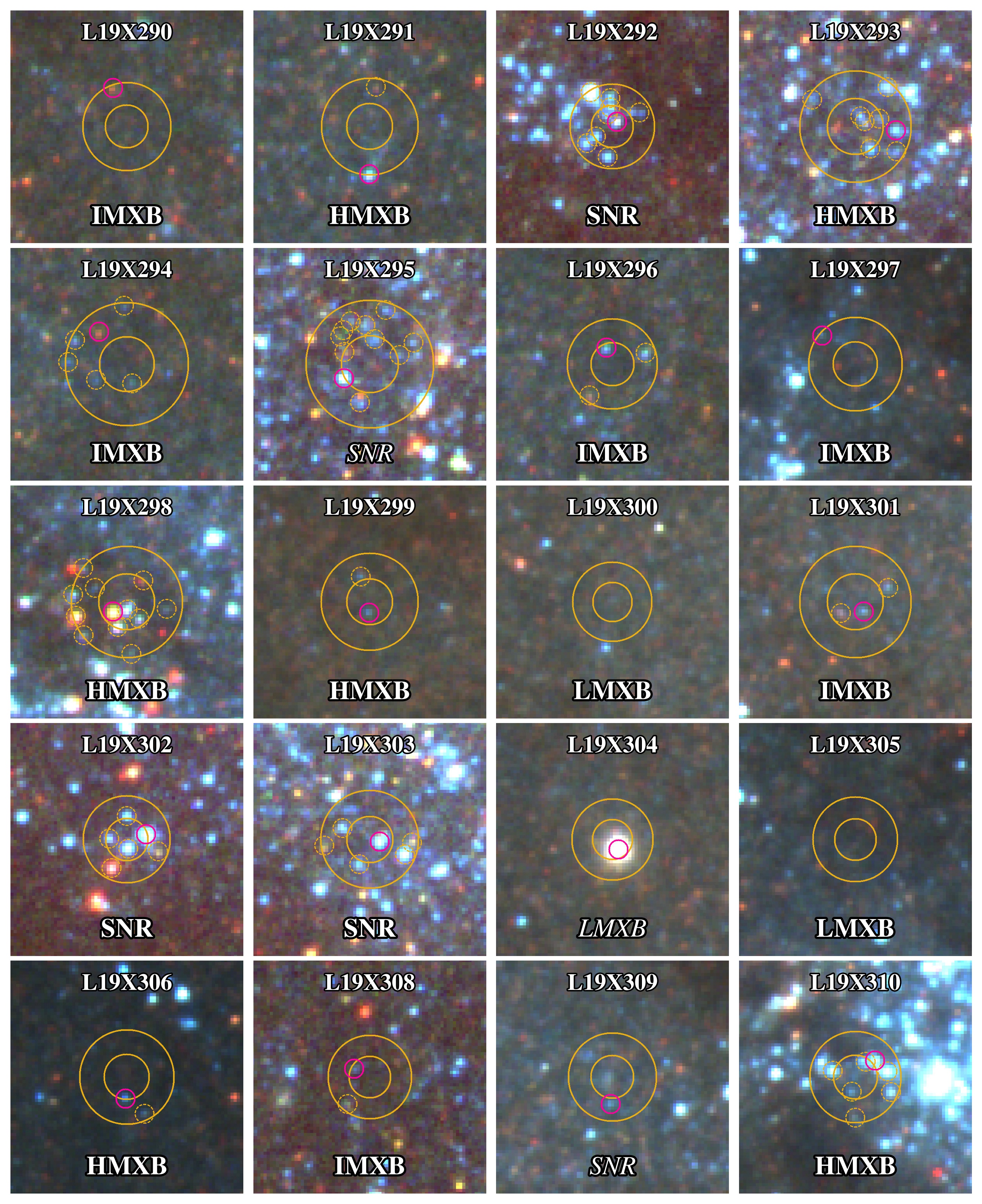}
\caption{\textit{(continued)}}
\end{figure*}

  \begin{figure*}
\ContinuedFloat
\includegraphics[width=\linewidth]{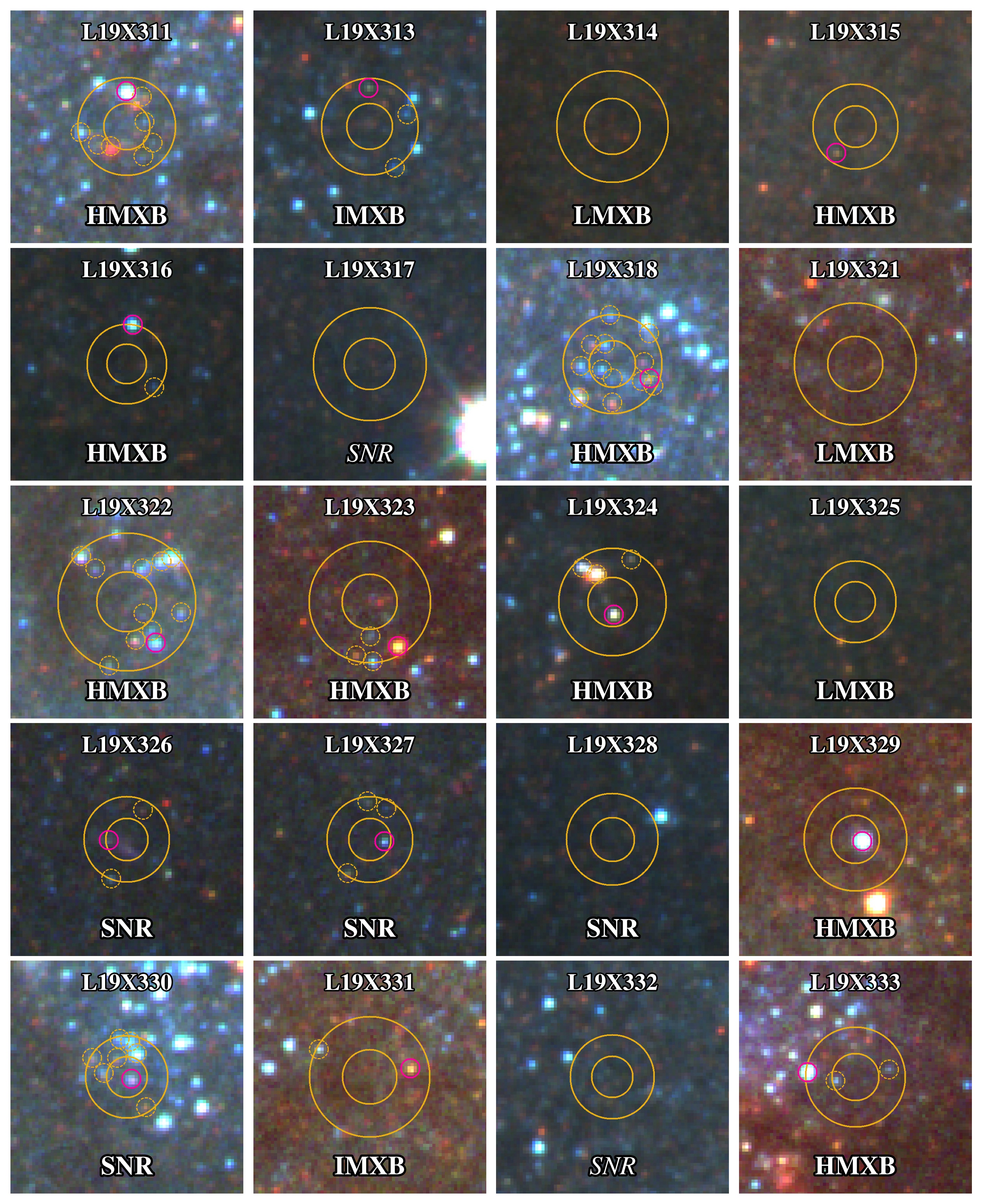}
\caption{\textit{(continued)}}
\end{figure*}

  \begin{figure*}
\ContinuedFloat
\includegraphics[width=\linewidth]{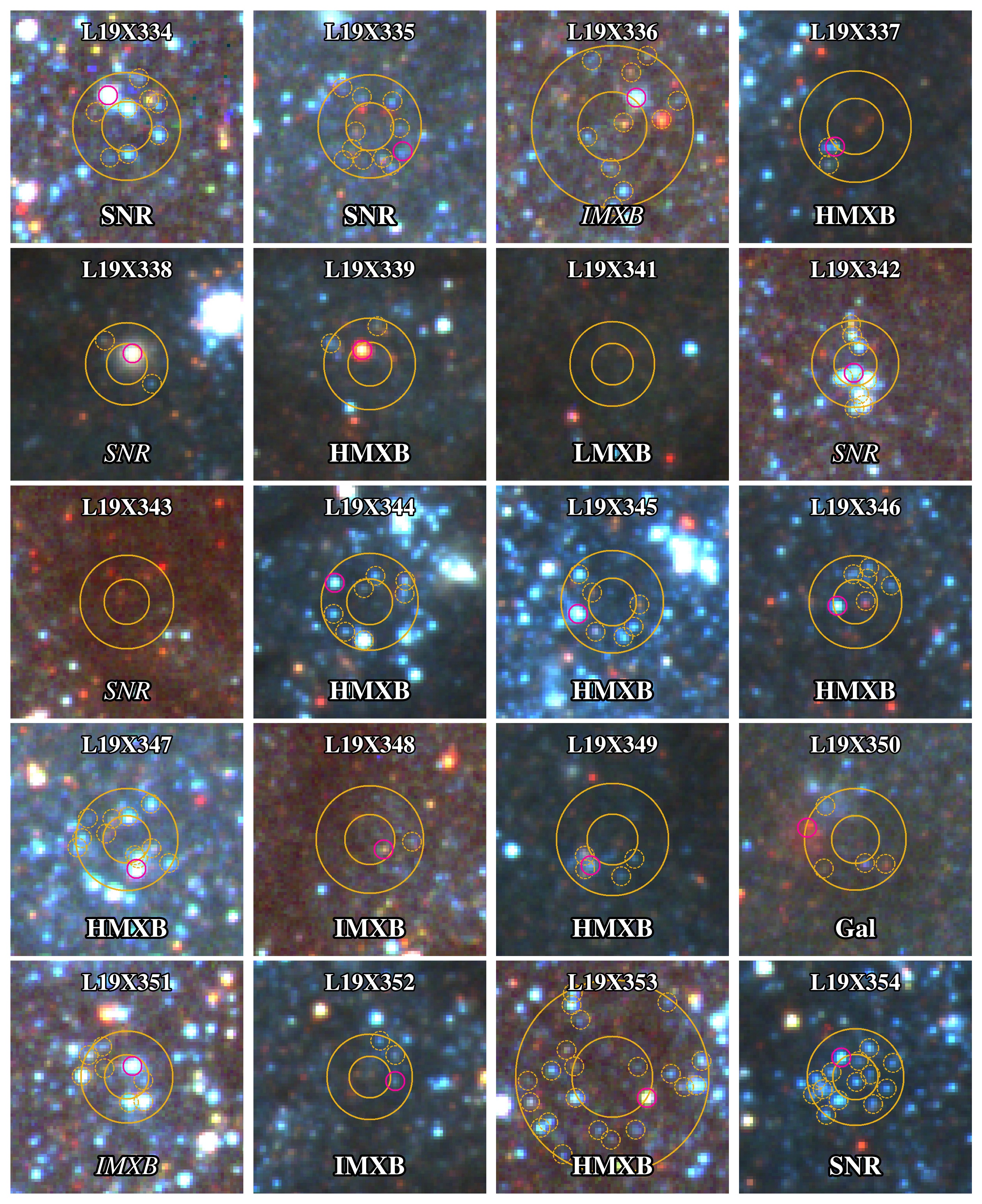}
\caption{\textit{(continued)}}
\end{figure*}

  \begin{figure*}
\ContinuedFloat
\includegraphics[width=\linewidth]{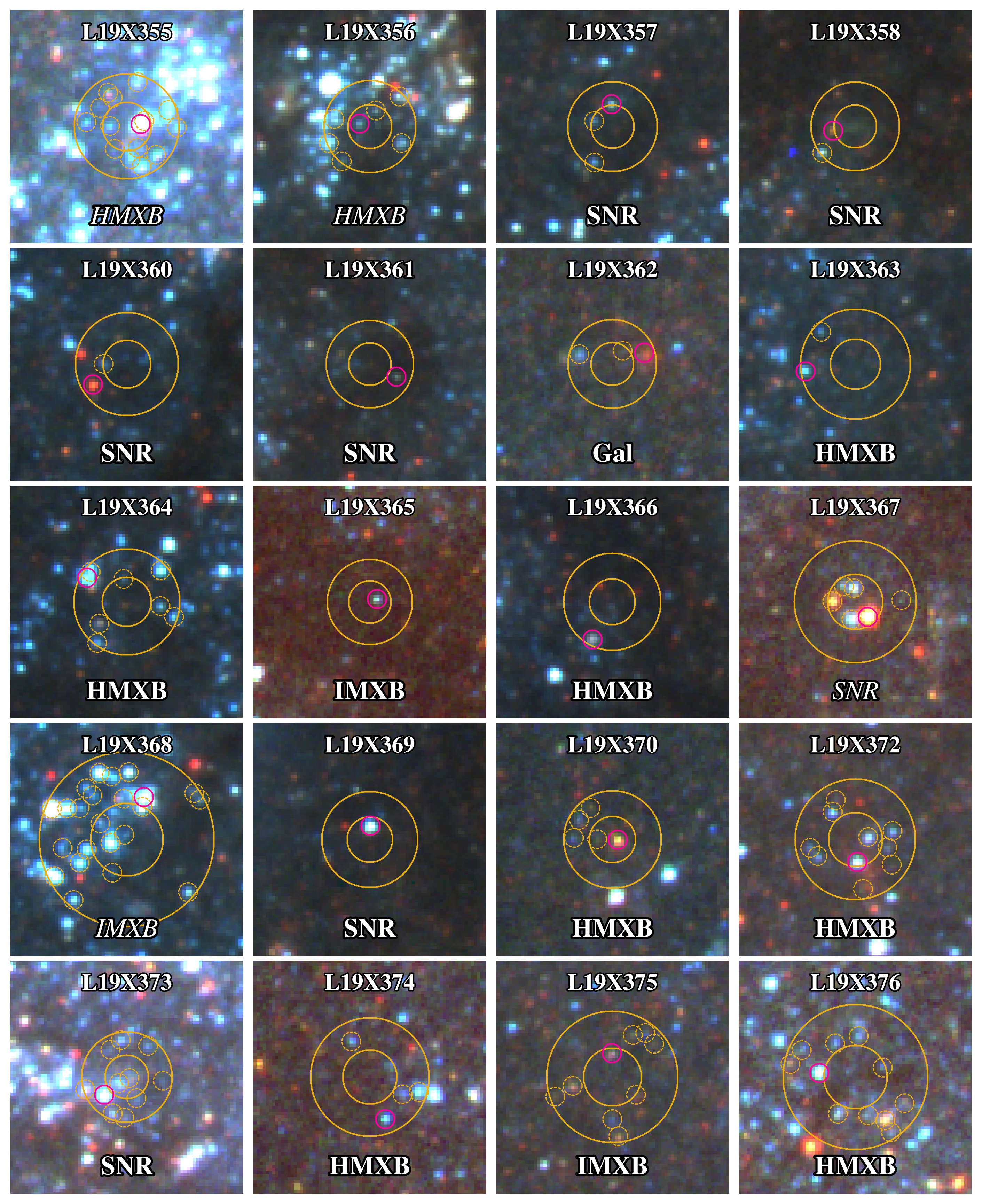}
\caption{\textit{(continued)}}
\end{figure*}

\begin{figure*}
\ContinuedFloat
\includegraphics[width=\linewidth]{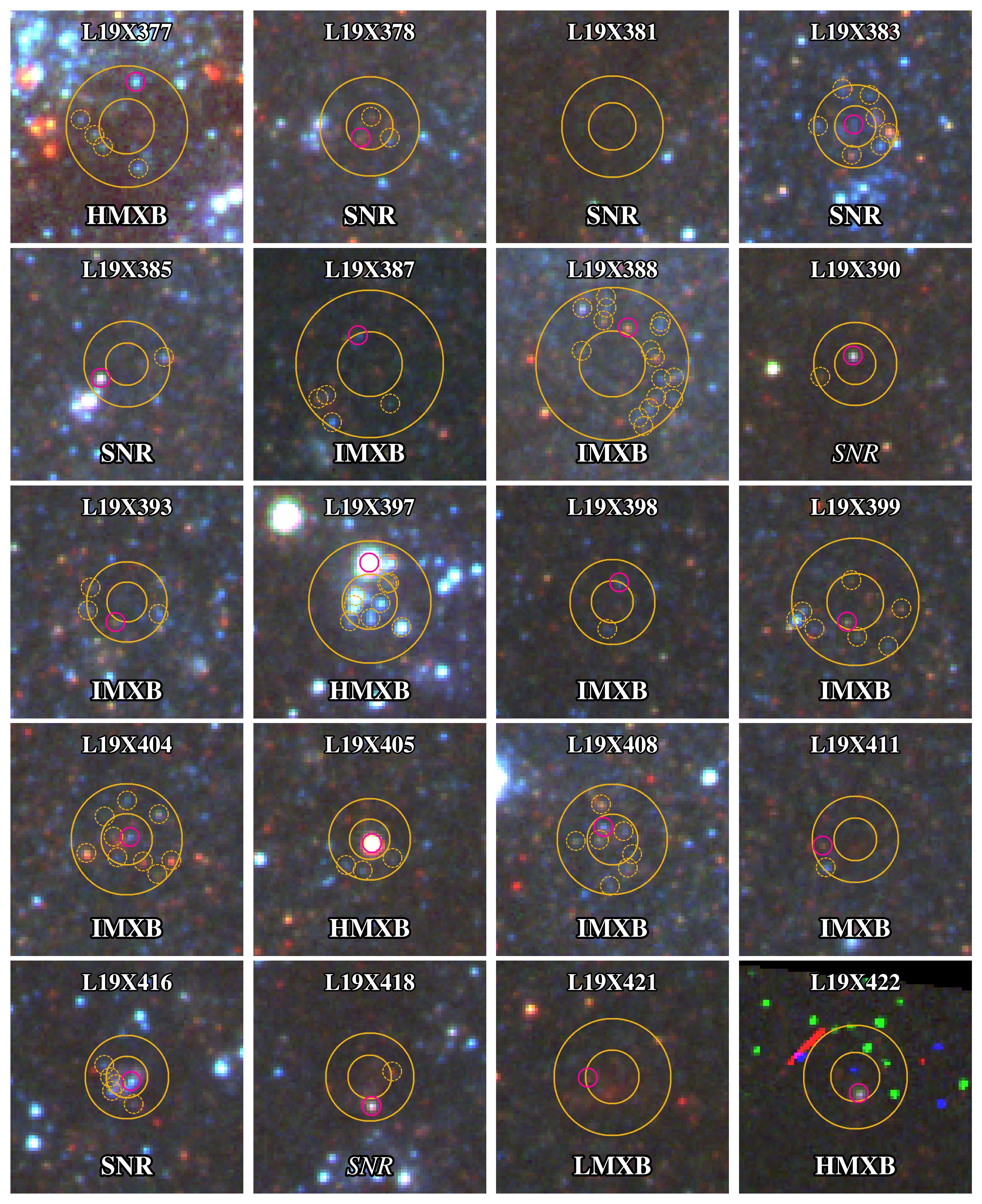}
\caption{\textit{(continued)}}
\end{figure*}

\begin{figure*}
\ContinuedFloat
\includegraphics[width=\linewidth]{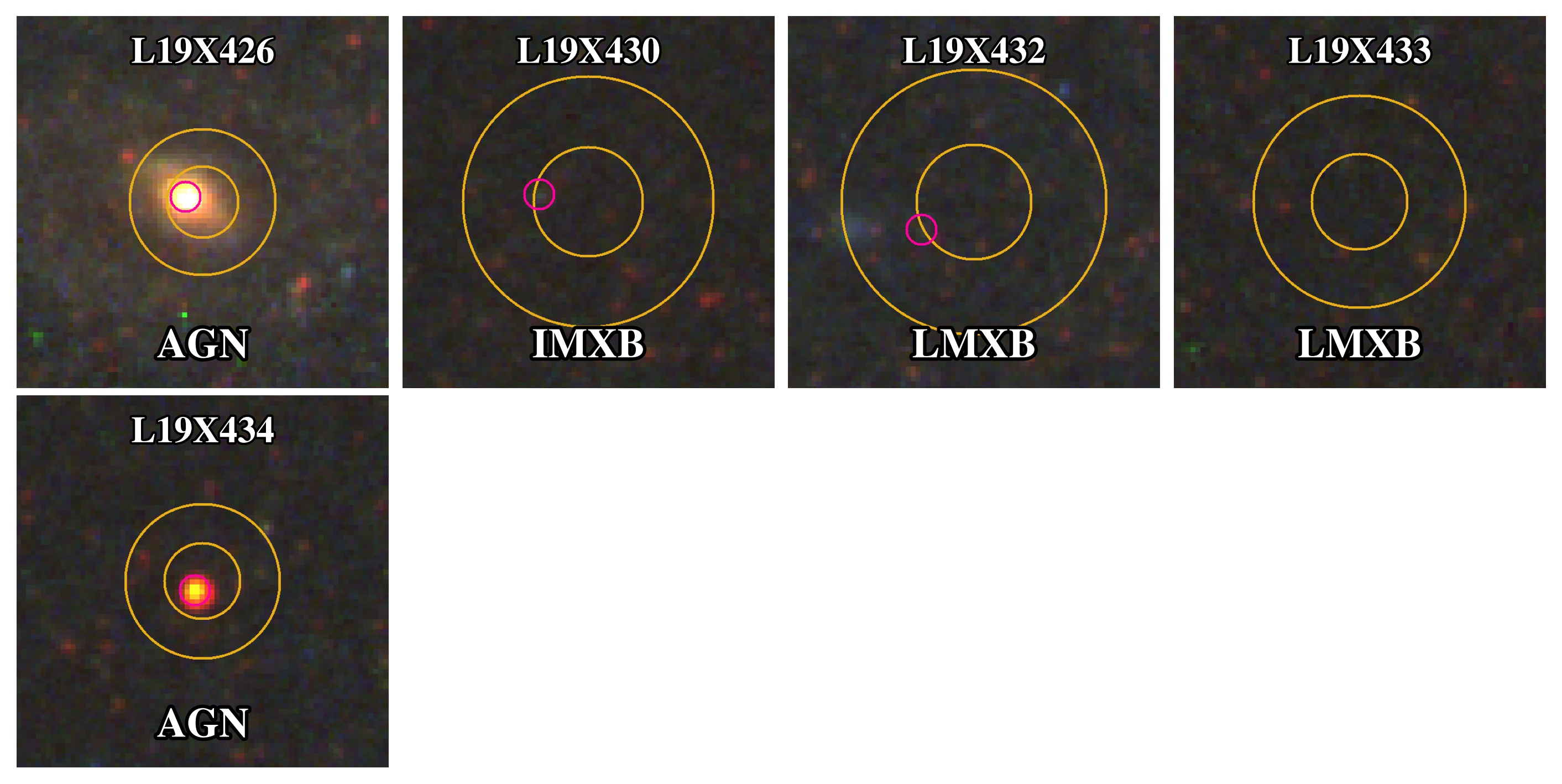}
\caption{\textit{(continued)}}
\end{figure*}

\end{document}